\documentclass{jfm}
\usepackage[utf8]{inputenc}
\usepackage{amsmath}
\usepackage{amssymb}
\usepackage{graphicx}
\usepackage{siunitx}
\usepackage{natbib}
\usepackage{xcolor}
\usepackage{bm}
\usepackage{comment}

\usepackage{hyperref}

\newcommand{\Jext}{J_\mathrm{ext}}
\newcommand{\Pext}{P_\mathrm{ext}}

\newcommand{\Ji}{J_\mathrm{i}}
\newcommand{\Jf}{J_\mathrm{f}}

\newcommand{\reffig}[1]{Fig.~\ref{#1}}
\newcommand{\figfolder}[1]{figures/}

\title{Transient flows and migration in granular suspensions: key role of Reynolds-like dilatancy}

\author{S. Athani\aff{1}
  \corresp{\email{shivakumar.athani@univ-grenoble-alpes.fr}},
  B. Metzger\aff{2},
  Y. Forterre\aff{2},
 \and  R. Mari\aff{1}}

\affiliation{\aff{1}Universit\'e Grenoble--Alpes, CNRS, Laboratoire Interdisciplinaire de Physique (LIPhy), 38000 Grenoble, France
\aff{2}Aix  Marseille  Univ.,  CNRS,  IUSTI,  13453  Marseille,  France}

\begin{document}

\maketitle

\begin{abstract}

We investigate the transient dynamics of a sheared suspension of neutrally buoyant particles under pressure-imposed conditions, subject to a sudden change in shear rate or external pressure. Discrete Element Method simulations show that, depending on the flow parameters (particle and system size, initial volume fraction),  the early stress response of the suspension may strongly differ from the prediction of the Suspension Balance Model based on the steady-state rheology. We show that a two-phase model incorporating the Reynolds-like dilatancy law of \cite{pailha_two-phase_2009}, which prescribes the dilation rate of the suspension over a strain scale $\gamma_0$, quantitatively captures the suspension dilation/compaction over the whole range of parameters investigated. Together with the Darcy flow induced by the pore pressure gradient during dilation or compaction, this Reynolds-like dilatancy implies that the early stress response of the suspension is nonlocal, with a nonlocal length scale $\ell$ which scales with the particle size and diverges algebraically at jamming. In regions affected by $\ell$, the stress level is fixed, not by the steady-state rheology, but by the Darcy fluid pressure gradient resulting from the dilation/compaction rate. Our results extend the validity of the Reynolds-like dilatancy flow rule, initially proposed for jammed suspensions, to flowing suspension below $\phi_\mathrm{c}$, thereby providing a unified framework to describe dilation and shear-induced migration. They pave the way for  understanding more complex unsteady flows of dense suspensions, such as impacts, transient avalanches or the impulsive response of shear-thickening suspensions.

\end{abstract}

\section{Introduction}

Granular suspensions consisting of rigid particles and fluids are ubiquitous in nature (landslides including debris flow, mud flow and submarine avalanches) and industrial applications (among others manufacturing and handling of concrete, drilling slurries or molten chocolate)~\citep{guazzelli2018rheology}.  In many flow situations, the particle volume fraction of the suspension is not constant but can evolve both in space and time. This is observed for instance in volume-imposed configurations with neutrally-buoyant particles, when shear-induced migration occurs as a result of an in inhomogeneous flow field, such as in pipe or large gap Couette flows~\citep{morris1999curvilinear}. It is also observed in pressure imposed configurations, for instance in gravity driven flows with non-buoyant particles, when an immersed granular avalanches dilates in order to flow~\citep{iversonAcuteSensitivityLandslide2000}. For both of these situations, the flow involves local changes of the volume fraction. Upon dilation, pore space is created in between the particles, which has to be filled by the incompressible fluid. The resulting relative motion between the fluid and solid phases creates a drag on the particles which transiently increases the particle pressure. A similar scenario, but of opposite sign, occurs upon compaction: this time fluid is expelled out of the contracting pores, which transiently decreases the particle pressure. 

Depending on the scientific community and the flow regime considered, this two-phase flow coupling has been apprehended in two very different ways.  In the suspension community, transient changes of the particle volume fraction below $\phi_\mathrm{c}$ (the critical volume fraction at which the suspension jams), are usually described using the so-called Suspension Balance Model (SBM)\citep{nottPressuredrivenFlowSuspensions1994,morrisPressuredrivenFlowSuspension1998,morris1999curvilinear,nottSuspensionBalanceModel2011}. In this framework, particle migration is driven by particle stress gradients arising in inhomogeneous flow field.  SBM is built on phase averaged momentum and mass conservation laws~\citep{jacksonLocallyAveragedEquations1997} and two closures: an interphase coupling and a constitutive law for the particle phase stress. In the Stokes regime, the interphase coupling is usually written as a drag proportional to the relative phase velocities. The standard choice for the constitutive law is the steady-state rheological flow rules for the shear and normal particle stresses, which have now been well characterized and tested~\citep{guazzelli2018rheology}. These constitutive laws can be equivalently expressed either as ``volume-imposed'' or ``pressure-imposed'' flow rules.  In volume-imposed configurations, dimensional analysis requires that, in steady state, stresses are viscous and proportional to $\eta_\mathrm{f}\dot\gamma$, where $\eta_\mathrm{f}$ is the fluid viscosity and $\dot{\gamma}$ the shear rate. The constitutive laws thus resume to the volume fraction dependence of the shear $\eta_s(\phi)$ and normal $\eta_n(\phi)$ viscosities, where $\phi$ is the particle volume fraction. In pressure-imposed configurations, the dimensionless number controlling the flow is the viscous number $J \equiv \eta_\mathrm{f}\dot\gamma/P_\mathrm{p}$, where $P_\mathrm{p}$ is the particle pressure. The constitutive laws are then provided by the relations $\mu(J)$ and $\phi_\mathrm{SS}(J)$, where $\mu = \tau/P_\mathrm{p}$ is the suspension friction coefficient, with $\tau$ the applied shear stress. So far, SBM has been tested and found to be in reasonable agreement with experimental results on particle migration in wide-gap Couette~\citep{morris1999curvilinear,sarabian2019fully}, in pipe flows of concentrated suspensions~\citep{snookDynamicsShearinducedMigration2016}, or during resuspension ~\citep{acrivos1993shear,d2021viscous,saint2019x}.

In the granular and soil mechanics communities, volumetric changes of the particle phase are usually studied for packings initially at rest, and prepared very close to, or even beyond $\phi_\mathrm{c}$. A central concept in this field is the Reynolds dilatancy~\citep{reynolds1885lvii,wood1990soil}, which stipulates that, in order to deform, an initial dense (resp. loose) packing must dilate (resp. compact) toward its critical state of volume fraction $\phi_\mathrm{c}$ in the quasistatic regime. In the presence of an interstitial fluid, this change in volume fraction leads to Darcy back-flow (or negative/positive pore pressure) whose drag on the particles sets the particle stress. This Darcy-Reynolds coupling between granular dilatancy and Darcy back-flow is essential for predicting the behavior of soils under drained or undrained conditions~\citep{wood1990soil}. It was also shown to have major consequences for the transient flow of granular suspensions close to jamming, such as for the onset of debris flows and submarine avalanches~\citep{iversonAcuteSensitivityLandslide2000,pailhaInitiationUnderwaterGranular2008,rondon2011granular,topin2012collapse,bougouin2018granular,montella2021two}, during impacts~\citep{jerome2016unifying}, silo discharge~\citep{kulkarni2010particle},  or when shearing dense clouds of particles~\citep{metzger2012clouds}.   To model this feedback,  \cite{pailha_two-phase_2009} proposed a dilatancy dynamics for $\phi$, which is essentially a relaxation of $\phi$ towards its steady-state law $\phi_\mathrm{SS}(J)$ on a typical strain scale  $\gamma_0\sim \mathcal{O}(1)$, extending previous Reynolds dilatancy laws proposed for dry granular media in the quasi-static regime~\cite{rouxTextureDependentRigidPlasticBehavior1998}.

Interestingly, the theory based on Reynolds dilatancy and the Suspension Balance Model describe the same type of interplay between dilation (compaction) of the granular skeleton and fluid interphase drag forces. However, this coupling is treated in very different ways. In the Suspension Balance Model, the particle stress field is set by the steady-state rheological flow rules and the migration rate adapts to satisfy the force balance between this stress field and the interphase drag force. Conversely, in the theory based on Reynolds dilatancy, the rate of dilation is geometrically imposed by a ``dilatancy angle" (the distance between the actual volume fraction to $\phi_\mathrm{SS}(J)$). In this case, it is the particle stress field that adapts to this transient kinematically-constrained evolution.  This fundamental difference raises important questions. A particularly pregnant issue is the origin of the stress levels observed during transient flows: Are the typical pore and particle pressures set by the steady-state rheology, or by the Darcy back-flow induced by a geometrically constrained dilation (compaction) of the granular phase? Another question is whether the concepts of Reynolds dilatancy and shear-induced migration could be reconciled and described in a unified framework?  Clearly, the Suspension Balance Model cannot be used for very compact granular layers, as above $\phi_\mathrm{c}$, the steady-state rheological flow rules are not defined. However, whether the Reynolds dilatancy concept applies to describe migration for systems below $\phi_\mathrm{c}$ has never been tested. 

\begin{figure}
\centering
  \includegraphics[width=\columnwidth]{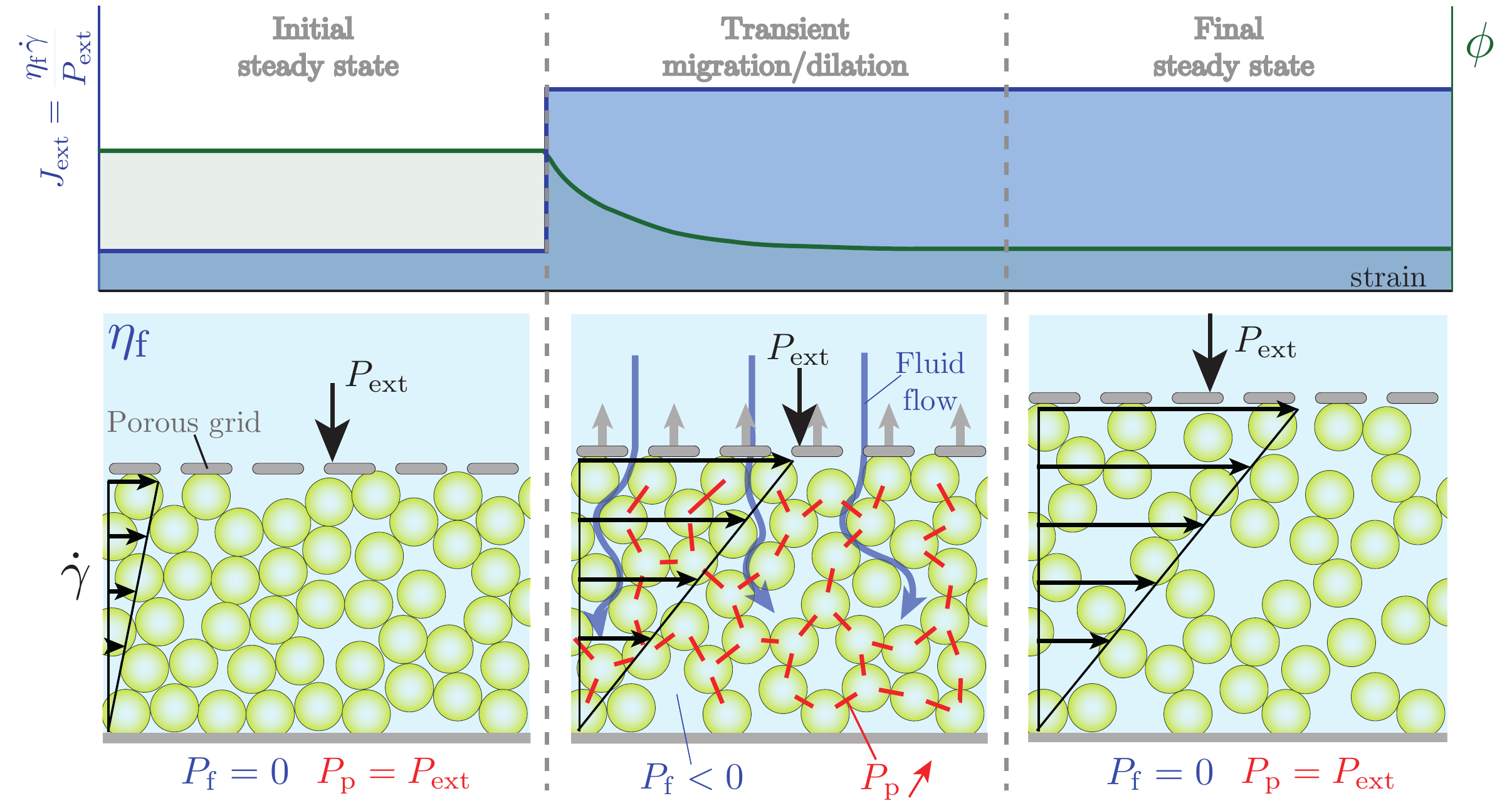}
\caption{\emph{Gedanken} experiment. A suspension of neutrally buoyant particles is homogeneously sheared under an imposed external pressure $P_{\rm{ext}}$ and shear rate $\dot{\gamma}$. A step change in $J_{\rm{ext}}=\eta_{\rm{f}} \dot{\gamma}/P_{\rm{ext}}$ is then applied by suddenly varying $P_{\rm{ext}}$ or $\dot{\gamma}$. Upon the step increase in $J_{\rm{ext}}$ shown in the Figure, the particle skeleton dilates and yields: a transient fluid Darcy back-flow (blue arrows), a transient decrease of pore pressure $P_{\rm{f}}$ and a transient increase of particle stress $P_\mathrm{p}$. The red segments highlight the increase of the particle stress.}
\label{system1}  
\end{figure}

In this article, we address the above questions by performing Discrete Element Method (DEM) simulations in a canonical configuration: a neutrally buoyant sheared suspension under pressure-imposed conditions, subject to a sudden change in shear rate or external pressure (i.e. a step change in $J_{\rm{ext}}=\eta_{\rm{f}} \dot{\gamma}/P_{\rm{ext}}$, see Figure \ref{system1}).  This simple configuration allows us a detailed characterization of the dilation (compaction) dynamics, while varying all the relevant control parameters (initial and final $J_{\rm{ext}}$, ratio of system size to particle radius $H/a$). We then compare these numerical results to the two different continuum models described above: the Suspension Balance Model based on the steady-state rheological flow rules (from now on simply called \emph{steady-state rheology model}), and the two phase model coupling the Darcy drag force and the Reynolds-like dilatancy closure of~\cite{pailha_two-phase_2009} (from now on called \emph{Darcy-Reynolds model}, following~\cite{jerome2016unifying}). 

We find that, while at large strains both continuum models give similar results in agreement with DEM, at small strains, only the Darcy-Reynolds model can quantitatively capture the DEM results over the whole range of control parameters. The steady-state approximation is not only quantitatively poorer, it also provides qualitatively wrong predictions. Specifically, we show that the Darcy-Reynolds model induces non-locality in the stress upon dilation (compaction), on a typical length scale $\ell$ proportional to the grain size and diverging at the jamming transition, which the steady-state rheology model is completely oblivious to.  This nonlocality implies that the transient excess of macroscopic stresses observed during dilation is size-dependent. Systems whose size $H$ is smaller than $\ell$ show smaller level of stresses than expected from the steady-state rheology. In this case, stresses are set by the Darcy back-flow induced by the volumetric strain. Conversely, systems much larger than $\ell$ show larger stresses, well predicted by the steady constitutive law. Overall, our results highlights that during transients, stress levels in granular suspensions, even below $\phi_\mathrm{c}$, are controlled by a Reynolds-like dilation (compaction) of the granular phase and not by the steady-state rheological flow rules.

 The manuscript is organized as follows: After presenting the numerical set-up in \S\,\ref{sec:numerics}, the DEM results are reported in \S\,\ref{DEMresults}. We then detail both continuum models in \S\,\ref{sec:modelling} and compare them to DEM results in \S\,\ref{comparison}. Conclusions are drawn in \S\,\ref{conclu}.


\section{\label{sec:numerics}Numerical Setup}

\begin{figure}
\centering
  \includegraphics[width=10cm]{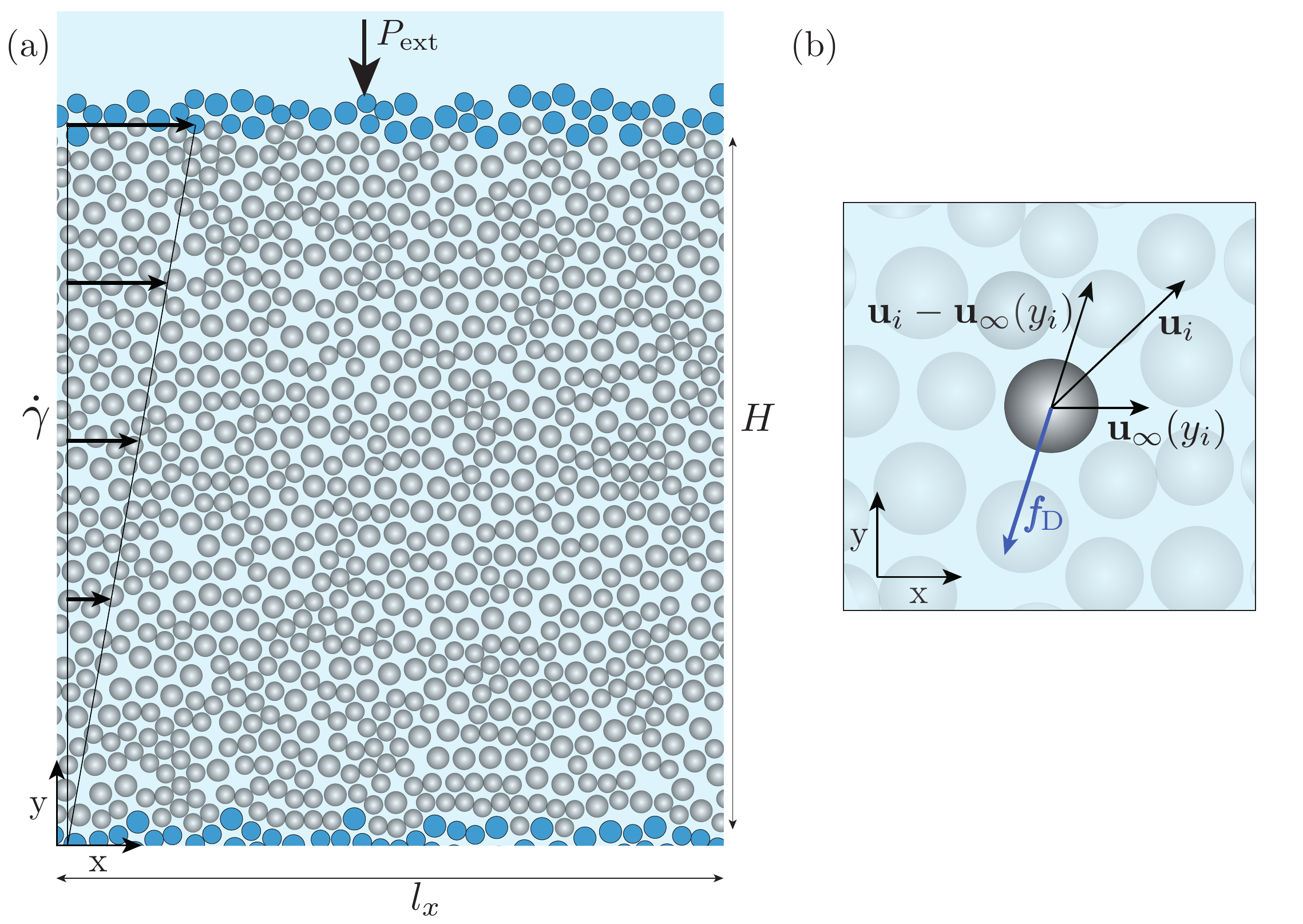}
\caption{a) Pressure-imposed DEM simulations. The system is composed of mobile particles (in grey) which are immersed in a suspending Newtonian fluid and placed between the top and bottom walls constructed with frozen particles (in blue). The control parameters are the external stress $P_\mathrm{ext}$ on the top-wall and the shear rate $\dot\gamma$ imposed to the entire system. In this pressure-imposed configuration,  the top wall can move vertically, allowing the volume fraction to adjust freely owing to the value of the imposed $J_{\rm{ext}}=\eta_\mathrm{f} \dot{\gamma}/P_{\rm{ext}}$. The fluid is not explicitly modeled but accounted for through lubrication and drag forces. b) The drag force $\bm{f}_\mathrm{D}$ exerted by the fluid on particle $i$ is proportional to the difference of between the local fluid and particle velocities. This force ensures that grains are advected by an homogeneous background shear flow in the $x$-direction, and models the pore pressure feedback (or interphase drag coupling) occurring during dilation (compaction) of the granular skeleton in the $y$-direction.}
\label{system2}       
\end{figure}

Pressure-imposed DEM simulations are developed  based on an established DEM code for volume-imposed simulations presented in detail in~\citep{mari2014shear}. As depicted in Fig.~\ref{system2} (a), we consider a monolayer of neutrally buoyant non-Brownian hard disks (in gray) immersed in a Newtonian fluid of viscosity $\eta_\mathrm{f}$, placed between two rigid top and bottom walls separated by $H$ and built out of frozen particles (in blue). The bottom wall is not permeable to the fluid nor to the particles, while the top wall is permeable to the fluid, but not to the particles. Furthermore, the bottom wall is fixed, whereas the top wall is free to move vertically, subject to an externally applied normal stress $P_\mathrm{ext}$, and follows a prescribed horizontal velocity $\dot\gamma H$ when the suspension is sheared at an imposed shear rate $\dot\gamma$. This corresponds to an applied viscous number $J_\mathrm{ext} \equiv \eta_\mathrm{f} \dot\gamma/P_\mathrm{ext}$. This setup is, in spirit, the two-dimensional equivalent of the pressure-imposed rheometer of \cite{boyer2011unifying}. We choose a monolayer setup over a three-dimensional one to be able to reach reasonably large linear extensions, without simulating a prohibitively large number of particles. The particle size distribution is bidisperse, with a size ratio $1.4$, mixed in equal volume. We use periodic boundary conditions along the horizontal $x$-direction, and call $l_x$ the system length in this direction. The solid fraction of the system is then $\phi \equiv N\pi a^2/(l_x H)$, with $a$ the mean particle radius. We assume that the system is in the limit of vanishing Reynolds number, and neglect inertial effects. In this limit, we thus follow the Stokesian Dynamics approach and do not explicitly simulate the fluid phase, which only comes in as hydrodynamic interactions acting on the particles~\citep{bradyStokesianDynamics1988}. The dynamics being overdamped, the equation of motion consists of mechanical force and torque balance on every particle.

The forces acting on bulk particles are contact forces, with a Coulomb frictional model with friction coefficient $\mu_\mathrm{p} = 0.5$, implemented in a standard Cundall-Strack manner~\citep{cundallDiscreteNumericalModel1979}, and hydrodynamic forces. The particle stiffness is chosen such that the particle overlap is always smaller than $\SI{2}{\percent}$ of their diameter. We neglect long-range hydrodynamic interactions, which are screened in a dense system, and consider only the short-ranged pairwise lubrication forces~\citep{mari2014shear}. Finally, a Stokes drag is applied to particles, and is assumed proportional to the relative velocity between a particle and the background fluid velocity evaluated at the center of the particle, $\bm{u}_\infty(y) = (\dot\gamma y, 0)$ if the particle is at height $y$: for particle $i$ with radius $a_i$ and velocity $\bm{u}_i$, the drag force is $\bm{f}_{\rm{D}}=-6\pi \eta_\mathrm{f} a_i (\bm{u}_i - \bm{u}_\infty(y_i))$, see Figure \ref{system2} (b).  This force plays two important roles: first it ensures the granular suspension is advected by an homogeneous background shear flow in the $x$-direction.  Second, it is used to model, in the simplest possible way, the pore pressure feedback (or interphase drag coupling) occurring during dilation (compaction) of the granular phase in the $y$-direction. Note that two important simplifications are made. First, we neglect the vertical fluid velocity component that should arise to satisfy the suspension incompressibility during dilation (compaction). Second, we do not explicitly simulate the fluid flow between pores, nor do we use standard mesoscopic models as the empirical hindered settling function proposed by~\citet{zaki1954sedimentation}, which scales as $(1-\phi)^\alpha$, with $\alpha>0$~\citep{davisSedimentationNoncolloidalParticles1985,morris1999curvilinear,snookDynamicsShearinducedMigration2016}.

This choice of interaction between the solid and fluid phases, through a simple Stokes drag force, retains the minimal physics needed to describe the transient dynamics considered here: a drag force proportional to the difference of vertical velocities between fluid and particle phase. Such a choice will prove trivial to model when it comes to the comparison of our DEM results with continuum modeling. This is an important aspect of our approach, as it will allow a much finer testing of the constitutive laws. Indeed, instead of having to attribute discrepancies between the theoretical model and the numerical simulations to one of the two closures (the interphase drag and the constitutive law), which can rapidly prove impossible to disentangle, we know here by construction that the observed discrepancies are coming from inaccurate constitutive modeling only. The downside to this modeling choice is that we will not achieve quantitative agreement with experimental data, but this was already prevented by our choice of a monolayer.

Forces acting on the top wall are the imposed external force $P_\mathrm{ext} l_x$, the contact and lubrication forces coming from interaction with the bulk particles, and a viscous drag proportional to the relative motion of the wall with respect to $\bm{u}_\infty(H)$, which reads $- \kappa  l_x (\bm{u}_\mathrm{wall} - \bm{u}_\infty(H))$, with $\bm{u}_\mathrm{wall} = (\dot\gamma H, \partial_t H)$ the velocity of the top wall and $\kappa$ the wall hydraulic resistance. Except when noted, we take $\kappa$ as if a Stokes drag was acting on each particle composing the wall, that is $\kappa = 6\pi \eta_\mathrm{f} N_\mathrm{wall}a/l_x$, with $N_\mathrm{wall}$ the number of particles in the top wall. We also performed simulations with $\kappa=0$ which yields similar results.
The vertical motion of the top wall is obtained by writing that the wall is force-free and its equation of motion is derived in Appendix~\ref{app:numerical_method}.

The simulation protocol is as follows: the suspension is first pre-sheared under constant pressure $P_{\rm{ext}}$ and constant shear rate $\dot{\gamma}$ so that it reaches an initial steady state characterized by the initial viscous number $J_\mathrm{i}$. 
We then suddenly impose a step change in $J_\mathrm{ext}$. This can equivalently be done by keeping $P_{\rm{ext}}$ constant while changing the imposed shear rate $\dot{\gamma}$ (as shown in Figure \ref{system1}), or by keeping $\dot{\gamma}$ constant while changing the imposed external pressure $P_{\rm{ext}}$. 
Before the system eventually reaches a final steady state characterized by the viscous number $J_\mathrm{f}$, we systematically investigate the transient evolution of the system by monitoring the evolution of the top wall vertical position $H$, the depth-averaged shear stress $\bar{\tau}$, and the volume fraction $\phi$ and particle vertical normal stress $P_\mathrm{p}$ profiles within the granular layer along the $y$-direction. Simulations are performed for both dilation ($J_\mathrm{f}>J_\mathrm{i}$) and compaction ($J_\mathrm{f}<J_\mathrm{i}$), over a wide range of viscous number $J_\mathrm{ext}\in [10^{-4}-10^{-1}]$, and for three systems sizes differing by their particle number ($N=1500, 1000$ and $600$) or corresponding vertical normalized height ($H/a=140, 68$ and $33$, respectively, when $J_{\rm{ext}}=10^{-4}$). 

\begin{figure}
\centering
  \includegraphics[width=0.8\columnwidth]{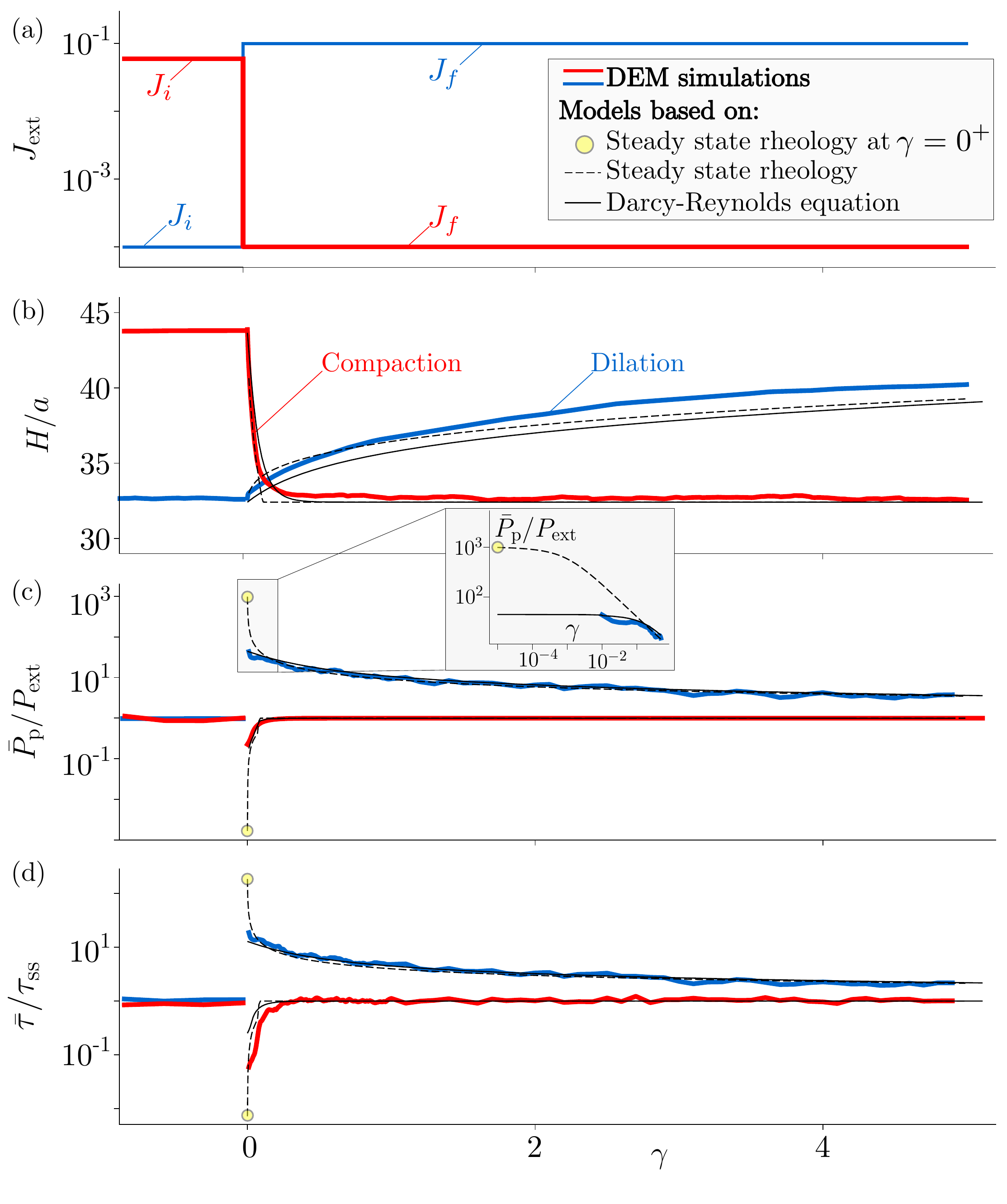}
\caption{Macroscopic response of the granular layer subject to a step increase (in blue) and a step decrease (in red) in $\Jext$ applied at $\gamma=0$. (a) Imposed $\Jext$, (b) evolution of the layer thickness height $H/a$, (c) normalized depth-averaged particle pressure $\bar{P}_\mathrm{p}/\Pext$, and (d) normalized depth-averaged shear stress $\bar{\tau}/\tau_{\rm{SS}}$ at the top wall, where $\tau_{\rm{SS}}=\mu(J_{\rm{ext}})P_{\rm{ext}}$ when the system is at steady state, versus strain $\gamma$. All results are averaged over $10$ different realizations and were obtained for the system size $H/a=33$.  Red and blue lines: DEM simulations, Dashed black lines: model based on steady-state rheology (Yellow circles: prediction at $\gamma=0^{+}$), Solid black lines: Darcy-Reynolds model.}
\label{fig:macro_pheno} 
\end{figure}

\section{DEM simulation results}
\label{DEMresults}

\subsection{Macroscopic phenomenology}

Figure \ref{fig:macro_pheno} shows the evolution of the granular layer subject to a step change in $\Jext$ for two cases of dilation ($J_\mathrm{f}>J_\mathrm{i}$, blue line) and compaction ($J_\mathrm{f}<J_\mathrm{i}$, red line). After the step, we observe that when $J_\mathrm{f}>J_\mathrm{i}$ (resp. $J_\mathrm{f}<J_\mathrm{i}$), the thickness of the granular layer $H$ increases (resp. decreases) as a result of the dilation (resp. compaction) of the granular skeleton.  This behavior is expected since the steady-state rheological rule $\phi_\mathrm{SS}(J)$ prescribes that the volume fraction of the suspension must decrease with its viscous number. Interestingly, the strain scale to reach the new steady state at $\Jf$ strongly differs for dilation and compaction. As seen in \reffig{fig:macro_pheno}a, the volumetric change occurs on a strain $\gamma$ of order 1 for compaction, while it takes more than 10 strain units to complete for dilation. This simply comes from the fact that dilation and compaction are dissymmetric as they occur under different external values of $J_{\rm ext}$.

\begin{figure}
\centering
  \includegraphics[width=\columnwidth]{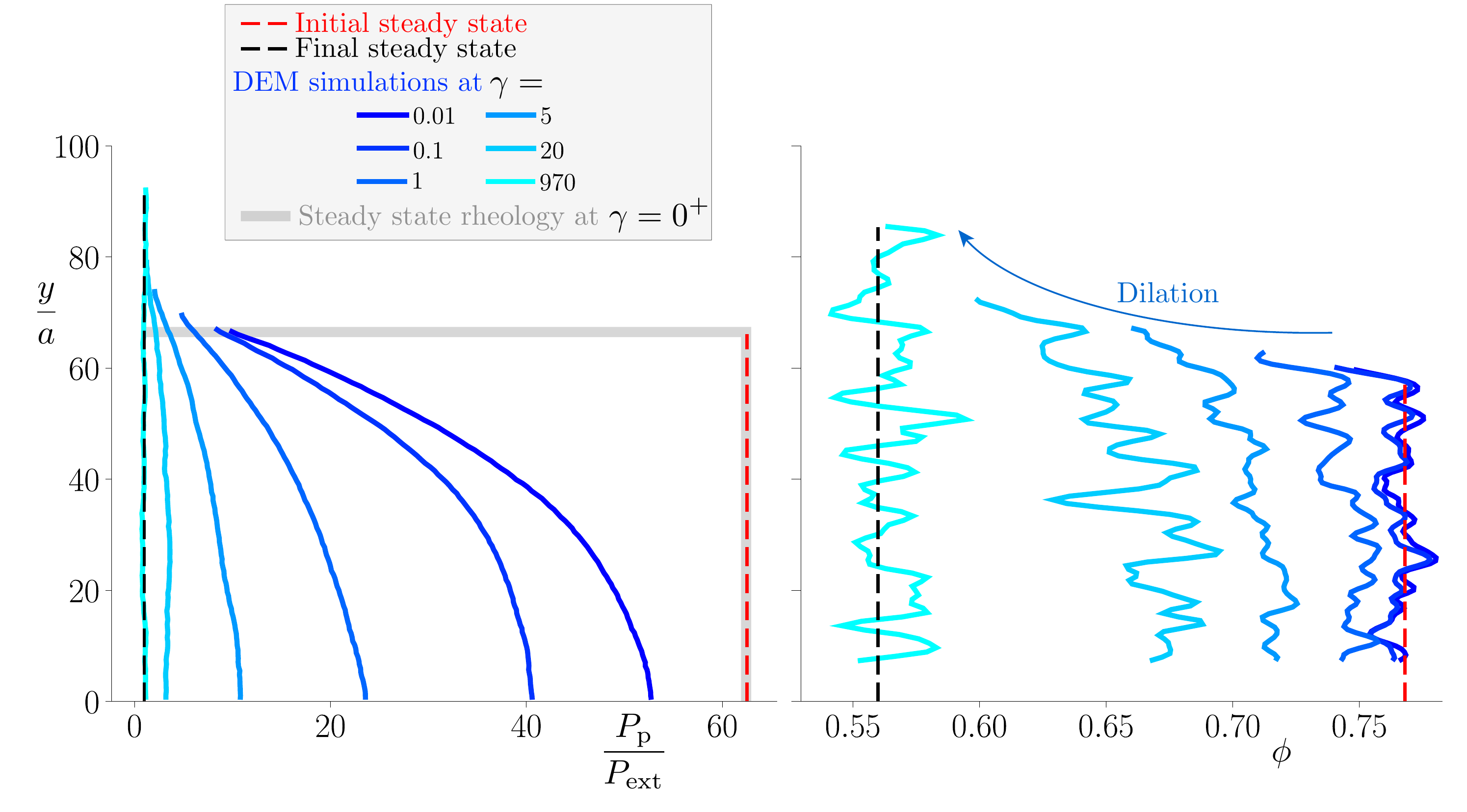}
\caption{Successive normalized particle pressure $P_{\rm p}/P_{\rm ext}$ and volume fraction $\phi$  profiles along the $y$-direction  after a step change in $\Jext$ (dilation case with $\Ji=\num{1.6e-3}$ and $\Jf=\num{e-1}$).}
\label{fig:pheno_profiles}       
\end{figure}

Remarkably, during the transient, both the mean particle pressure $\bar{P}_\mathrm{p}/\Pext$ (averaged over the whole layer thickness) and the depth-averaged shear stress $\bar{\tau}/\tau_{\rm{SS}}$ exhibit a sudden and large increase (resp. decrease) for dilation (resp. compaction), before relaxing towards their steady-state value. 
Understanding what sets the stress levels during these transients is precisely one of the main goal of the present study. We can readily compare the magnitude of these stress jumps, occurring right after the step, to the prediction of the steady-state rheology model. At $\gamma=0^{+}$, the particle packing fraction profile must be the same as that of the initial steady-state preparation. It is therefore homogeneous along the $y$-direction such that $\phi |_{\gamma=0^{+}}=\phi|_{\gamma=0^{-}}=\phi_\mathrm{SS}(J_\mathrm{i})$. Assuming the steady-state flow rule $\phi_\mathrm{SS}(J)$ applies at all times, the viscous number inside the granular layer just after the jump is thus also the same as that just before the jump, i.e. $J |_{\gamma=0^{+}}=J|_{\gamma=0^{-}}=J_\mathrm{i}$, even though the external imposed viscous number has changed to $J_{\rm{ext}}=J_\mathrm{f}$ . Since the shear rate is imposed to be homogeneous, the particle pressure profile must also remain homogenous and equal to $P_{\rm{p}}|_{\gamma=0^{+}}=(\eta_{\rm{f}}\dot{\gamma}/J)|_{\gamma=0^{+}}=\eta_{\rm{f}}\dot{\gamma}|_{\gamma=0^{+}}/J_\mathrm{i}$ yielding
\begin{equation}
\frac{P_{\rm{p}}}{P_{\rm{ext}}} \bigg |_{\gamma=0^{+}}=\frac{J_{\rm{f}}}{J_{\rm{i}}}.
\label{SSpredictionP_p}
\end{equation}
Similarly, one obtains $(\bar{\tau}/\tau_{\rm{SS}})|_{\gamma=0^{+}}=\mu(J_\mathrm{i})J_{\rm{f}}/\mu(J_\mathrm{f})J_{\rm{i}}$. The steady-state rheology model thus predicts that, right after the step, the magnitude of the particle stress jump is given by the ratio of applied $J_\mathrm{ext}$, after and before the step, as highlighted by the yellow circles in Figure \ref{fig:macro_pheno}. 
We find that this prediction overestimates by more than one order of magnitude the particle pressure obtained from the DEM simulations. 
The same discrepancy is observed for the shear stress overshoots, $\bar{\tau}/\tau_\mathrm{SS}$, which are largely overestimated by the steady-state rheology.

\begin{figure}
\centering
\includegraphics[width=\columnwidth]{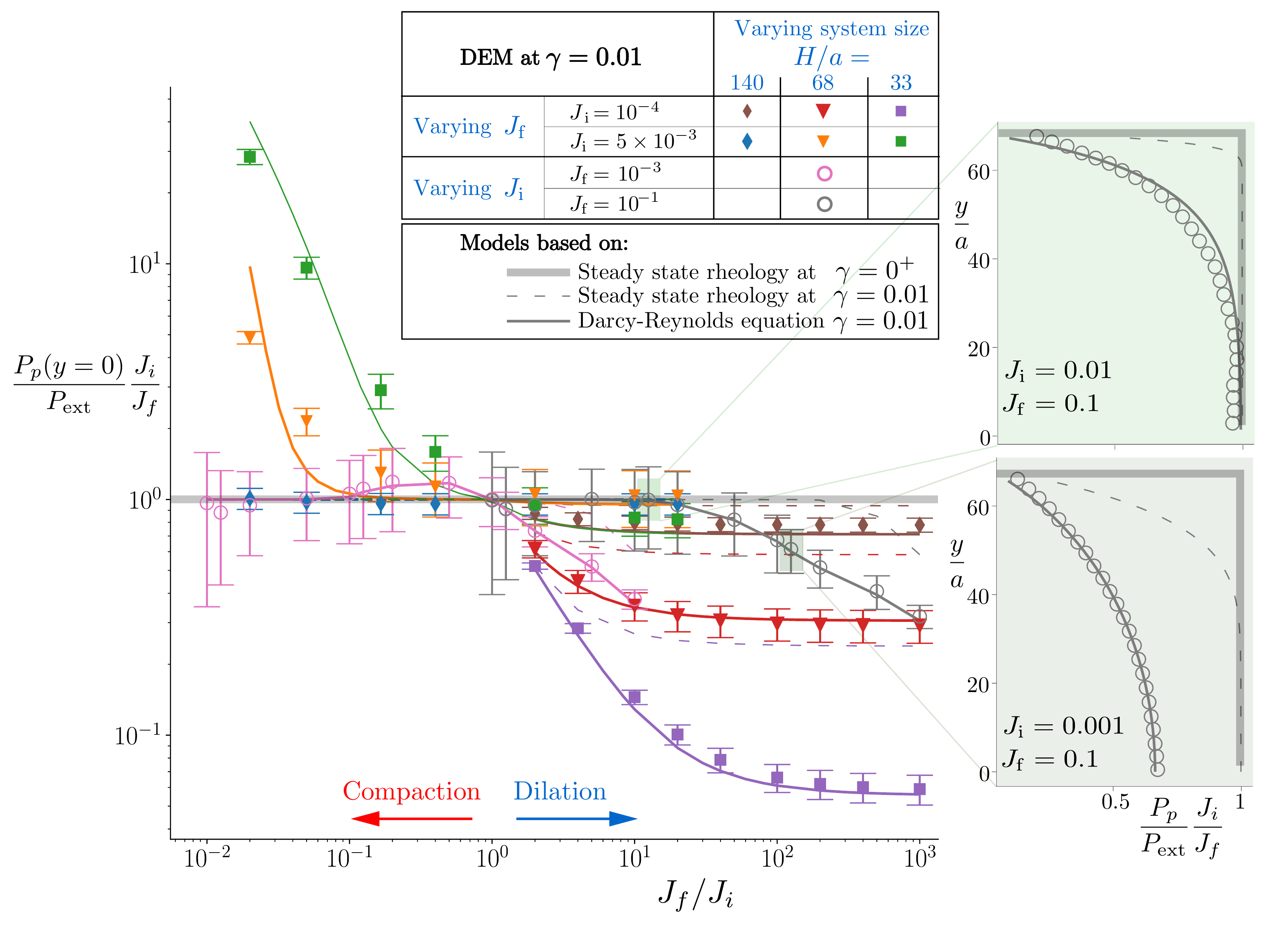}
\caption{Normalized particle pressure at the bottom of the cell $P_{\rm p}(y=0)J_{\rm i}/P_{\rm{ext}}J_{\rm f}$ just after the step change in $\Jext$ for various combination of $J_{\rm i}$ and $J_{\rm f}$ values, and different system sizes $H/a$. Markers:  DEM simulations evaluated at $\gamma= 0.01$. 
Thick grey line: prediction of the steady-state rheology at $\gamma= 0^{+}$,
Dashed lines: prediction of the steady-state rheology at $\gamma= 0.01$,
Solid lines: prediction of the Darcy-Reynolds model at $\gamma= 0.01$.
DEM simulation results are averaged over at least $5$ different realizations when $\Ji$ is fixed, and at least $40$ different realizations when $\Jf$ is fixed. Inserts: two examples of particle pressure profiles along the system height.}  
\label{fig:bottom_p_scatter}       
\end{figure}

\subsection{Transient evolution inside the bulk}

To further investigate the transient evolution of the granular layer, we now turn to local measurements of the particle pressure $P_\mathrm{p}(y)$ and solid fraction $\phi(y)$ profiles inside the bulk of the granular layer for a step increase in $\Jext$ (dilation case), see Figure \ref{fig:pheno_profiles}. After the step, the particle pressure $P_\mathrm{p}(y)$ and the volume fraction $\phi(y)$ progressively evolve from their initial (red dashed line) to final (black dashed lines) steady-state profiles. The striking observation is that, even at a strain as small as $\gamma=0.01$ after the step, the particle profile obtained from the DEM simulations (dark blue solid line) significantly departs from the prediction of the steady-state rheology model evaluated at $\gamma=0^{+}$: $P_{\rm{p}}(\gamma=0^{+},y)/P_{\rm{ext}}=\mathrm{constant}=J_{\rm{f}}/J_{\rm{i}}$ (grey solid line). The particle pressure exhibit smaller stress levels than expected and  a large gradient over the cell height, although the volume fraction profile is nearly uniform. These observations further evidence the inability of the steady-state rheology to describe the early stress state of the granular layer.


To characterize more systematically these differences between the prediction of the steady-state rheology and DEM results, we perform simulations for many combinations of the three control parameters of the system: $J_{\rm i}$, $J_{\rm f}$ and $H/a$. In Figure \ref{fig:bottom_p_scatter}, we report the normalized particle pressure at the bottom of the cell $P_{\rm p}(y=0)J_{\rm i}/P_{\rm{ext}}J_{\rm f}$, just after the step change in $J_{\rm ext}$, versus $J_{\rm f}/J_{\rm i}$. In this representation, the steady-state rheology prediction at $\gamma=0^{+}$ (Eq. \ref{SSpredictionP_p}) corresponds to the horizontal line $P_{\rm p}(y=0)J_{\rm i}/P_{\rm{ext}}J_{\rm f}=1$ (thick grey line). We find that many simulation data are in good agreement with this prediction, but equally many deviate from it, sometimes by more than an order of magnitude. We can also observe that the largest discrepancy occurs for small systems (purple and green squares in Figure \ref{fig:bottom_p_scatter}), while in large systems the pressure tends to be well predicted by the steady-state rheology model (brown and blue diamonds).
\\

The above results suggest that the steady-state rheology is in many cases unable to predict the early transient evolution of the stress levels in the granular suspension. Note that in practice, the earliest strain at which DEM results are reported in Fig. \ref{fig:pheno_profiles} and \ref{fig:bottom_p_scatter} is $\gamma=0.01$, since before that, DEM results may be dependent on the stiffness of the particles. One could thus argue that the discrepancy between DEM results and the steady-state rheology predictions arises from the fact they are not evaluated exactly at the same strains ($\gamma=0.01$ versus $\gamma=0^{+}$, respectively).  However, in what follows, by deriving the full transient evolution of the granular layer using both the steady-state and Darcy-Reynolds models, we show that this discrepancy arises from a more fundamental mechanism.

\section{\label{sec:modelling}Continuum models}

In this section, we recall the usual two-phase continuum description for suspensions~\citep{jacksonLocallyAveragedEquations1997},  expressed in the specific setup and approximation that we adopted in the DEM: a simplified expression for the interaction between the particle and fluid phase based on a Stokes drag, which neglects the fluid vertical counterflow induced by the dilation (compaction) of the particle phase. We then present two competing constitutive models to close this continuum description: one assuming steady rheology at all time (steady rheology model), the other taking into account a Reynolds-like dilatancy equation for the transient dynamics of the particle packing fraction (Darcy-Reynolds model).  

\subsection{Conservation laws}

The continuum model aims at describing the time evolution of the particle phase stress profile 
$P_\mathrm{p}(y)$ in the $y-$direction and solid fraction profile $\phi(y)$, and \emph{in fine} of the height $H$ of the top wall. It relies on the mass and momentum conservations and a constitutive relation relating the volume fraction $\phi$ with the particle stress $P_\mathrm{p}$ and the shear rate $\dot{\gamma}$. As our dilation problem is translation invariant in the $x-$horizontal direction, it is essentially a one-dimensional problem in the $y-$direction. We will call $u(y)$ the vertical velocity of the particle phase.
Mass conservation for the particle phase reads 
\begin{equation}
\partial_t \phi + \partial_y (u \phi) = 0.
\label{eq:mass}
\end{equation}

Momentum conservation, in the approximation of the Suspension Balance Model (SBM)~\citep{nottPressuredrivenFlowSuspensions1994,morrisPressuredrivenFlowSuspension1998,morris1999curvilinear,nottSuspensionBalanceModel2011},
balances the pressure gradient $\partial_y P_\mathrm{p}(y)$ with the local interphase drag, which is directly proportional to $u$ as we recall that the fluid vertical velocity is neglected\footnote{Note that the drag would still be proportional to $u$ even if we considered the fluid velocity, only the $\phi$ dependency of $R$ would be different.}
\begin{equation}
 -\partial_y P_\mathrm{p}(y) = \frac{\eta_\mathrm{f}\phi R(\phi)}{a^2} u(y),
 \label{eq:SBM}
\end{equation}
with $R(\phi)$ the hydrodynamic resistance of the particle matrix, which is dimensionless in 3D. 
This expression is adapted to our 2D DEM simulations with a simple Stokes drag using $R(\phi)=6a$, where the particle pressure $P_\mathrm{p}$ is now a force per unit length. Note that the linear dependence in $\phi$ of the interphase drag differs from the SBM as usually presented in the literature. This is to remain consistent with our numerical setup in which the interphase drag (\ref{eq:SBM}) is borne from the Stokes drag on every particle, hence is linear in $\phi$, rather than from an actual pore flow, which dependence in $\phi$ is more complex.  Besides its adequacy for the comparison with our simulation results, this simplification, while affecting the model accuracy with respect  to the flow of an actual suspension (but so does our choice of a two-dimensional setup), does not impact the qualitative features of the model.

The boundary conditions for the problem are as follows.  We require vanishing vertical velocity at the bottom wall $u(0) = 0$, which from (\ref{eq:SBM}) yields $\partial_y P_\mathrm{p}(y=0)=0$. Moreover, force balance on the top wall prescribes
\begin{equation}
\Pext + \kappa \partial_t H - P_\mathrm{p}(H) = 0,
\end{equation}
with $\kappa$ the resistance of the top wall, and we impose that the velocity of the particle phase follows the top wall velocity in $y=H$, $u(H) = \partial_t H$. Similarly, to match our 2D DEM setup, we use $\kappa =  6\pi \eta_\mathrm{f} N_\mathrm{wall}a/l_x$, with $N_\mathrm{wall}$ the number of particles in the top wall. We can adimensionalize this model with $a$ the unit length, $1/\dot\gamma$ the unit time,  and $\eta_\mathrm{f} \dot \gamma$ ($\eta_\mathrm{f} \dot \gamma a$ in 2D) the unit stress. We denote by $\hat{X}$ the adimensionalized $X$, for any physical quantity $X$. Inserting momentum balance in the mass balance equation to eliminate $u$, we obtain
\begin{equation}
\partial_{\hat{t}} \phi - \partial_{\hat{y}} \left[ \hat{R}(\phi)^{-1} \partial_{\hat{y}} \hat{P_\mathrm{p}} \right] = 0,
\label{eq:adim_balances}
\end{equation}
with boundary conditions 
\begin{align}
0 & = \left.\partial_{\hat{y}} \hat{P_\mathrm{p}}\right|_{0} \\
0 & = \Jext^{-1} - \frac{\hat{\kappa}}{\phi(\hat{H}) \hat{R}(\phi(\hat{H}))} \left.\partial_{\hat{y}} \hat{P_\mathrm{p}}\right|_{H} - \hat{P_\mathrm{p}}(\hat{H}),
\label{eq:BC}
\end{align}
with $\hat{\kappa}=\kappa a/\eta$ and $\hat{R}=R(\phi)$ in 3D, and $\hat{\kappa}=\kappa /\eta$ and $\hat{R}=R(\phi)/a$ in 2D. 

As such, the model is not closed, as we need a further relation between the normal stress and the solid fraction, that is, a constitutive model. In the following, we will consider two of them, both reducing to the usual [$\mu(J)$,  $\phi_\mathrm{SS}(J)$] rheology in steady state, but differing in their transient behaviors.

\subsection{Steady-state rheology model}

In the first model, we assume, as it is usually done when solving a shear-induced migration problem ~\citep{morris1999curvilinear,snookDynamicsShearinducedMigration2016,sarabian2019fully}, that the steady-state rheology is valid at all times and can describe the state of the suspension even during the transient migration of the particle phase. This simply sets
\begin{equation}
	\phi = \phi_\mathrm{SS}(J),
	\label{eq:SS_model}
\end{equation}
with $J=1/\hat{P}_p$ and $\phi_\mathrm{SS}(J)$ behaving close to jamming as
\begin{equation}
	\phi_\mathrm{SS}(J) \approx \phi_\mathrm{c} - KJ^{\beta},
	\label{eq:phi_SS}
\end{equation}
which is equivalent to $\eta_n(\phi)=K^{1/\beta}(\phi_\mathrm{c}-\phi)^{-1/\beta}$ in a volume imposed formulation~\citep{degiuli2015}.
Inserting Eq.~\ref{eq:SS_model} in Eq.~\ref{eq:adim_balances}, we obtain that the volume fraction obeys the following diffusion equation
\begin{equation}
	\partial_{\hat{t}}  \phi - \partial_{\hat{y}} \left[ D(\phi) \partial_{\hat{y}}  \phi \right] = 0,
	\label{eq:SS_PDE}
\end{equation}
with a $\phi$-dependent diffusion coefficient $D(\phi) = K^{1/\beta}/(\hat{R}(\phi) \beta J^{1+\beta})$.  

\begin{figure}
\centering
\includegraphics[width=11.5cm]{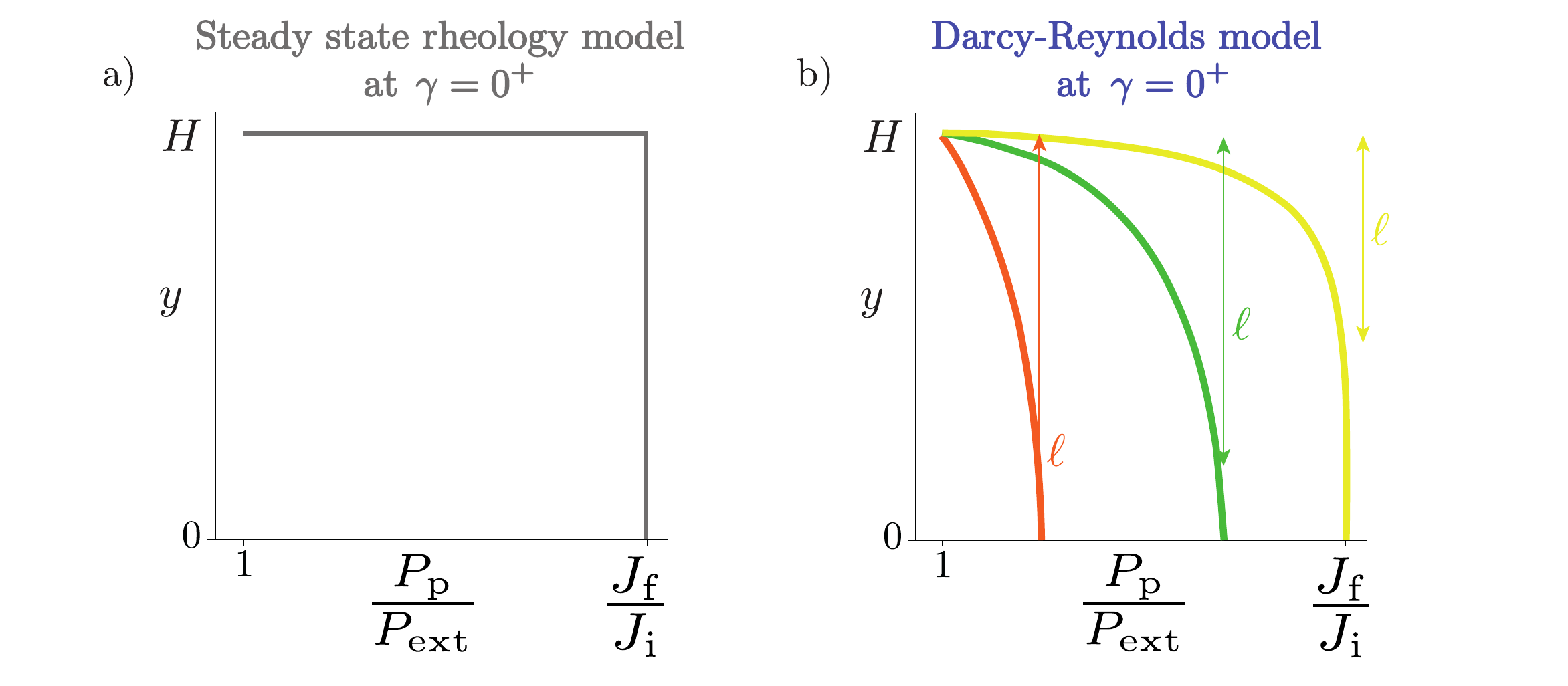}
\caption{Expected normalized particle pressure profile at $\gamma=0^+$ for a) the steady-state rheology model and b) the Darcy-Reynolds model for different values of $H/\ell$.}  
\label{fig:sketchmodel}       
\end{figure}

As for the temperature governed by the heat equation,  after a step change of external pressure on the upper wall or after a global change of shear rate, the particle stress remains strictly homogeneous at $\gamma = 0^+$ and exhibit a discontinuity at the upper wall boundary, see Fig. \ref{fig:sketchmodel} (a). The dilation/compaction process starts from the upper wall and gradually diffuses within the bulk of the granular layer, simply as heat would diffuse within the layer after a step change in temperature of the upper wall, with a $\phi$-dependent diffusion coefficient.

\subsection{Darcy-Reynolds model}

In the second model, we use the constitutive relation proposed by \cite{pailha_two-phase_2009}, which assumes that during transients the volume fraction $\phi$ relaxes towards its steady-state value $\phi_\mathrm{SS}(J)$ in a finite strain scale $\gamma_0$ as
\begin{equation}
	\partial_{\hat{t}} \phi = - \frac{1}{\gamma_0}\left[\phi - \phi_\mathrm{SS}(J)\right].
	\label{eq:PP_model}
\end{equation}
This simple closure was built as an extension at finite $J$ of previous Reynolds dilatancy laws proposed  for dry granular material~\citep{rouxTextureDependentRigidPlasticBehavior1998,rouxStatisticalApproachMechanical2002}, and to recover the  $\phi_\mathrm{SS}(J)$ rheology in steady state. It prescribes that the rate of dilation of the suspension is kinematically imposed by a ``dilatancy angle" propotional to the distance between the actual volume fraction $\phi$ and $\phi_\mathrm{SS}(J)$. The relaxation strain scale could  be $\phi$ dependent, but for the sake of simplicity we will consider it a material constant, as a fit parameter.

Crucially, this finite relaxation strain scale for dilation completely modifies the way the stress profile is set up.  In contrast to the diffusive dynamics for the steady-state rheology model (Eq.~\ref{eq:SS_PDE}),  inserting (\ref{eq:PP_model}) in (\ref{eq:adim_balances}), yields a second-order nonlinear ODE for the stress
\begin{equation}
	\partial_{\hat{y}}\left[\hat{R}(\phi)^{-1} \partial_{\hat{y}} \hat{P_\mathrm{p}}\right] + \gamma_0^{-1} \left[\phi - \phi_\mathrm{SS}(J)\right] = 0,
	\label{eq:PP_ODE}
\end{equation}
where  the time derivative has dropped. This fundamental difference has an important consequence regarding the early transient response during a change of $\Jext$ (or more generally to a change in boundary conditions).  In the ODE (Eq.~\ref{eq:PP_ODE}),
information can travel infinitely fast and instantaneous changes of the particle stress can occur at $\gamma = 0^+$ a finite distance from the wall. 
Said otherwise, in contrast with the steady-state rheology model, the Darcy-Reynolds model induces a non-local dynamics for the particle stress\footnote{While Eq.~\ref{eq:PP_ODE} formally looks similar to the stress dynamics postulated in non-local constitutive models~\citep{goyonSpatialCooperativitySoft2008,kamrinNonlocalConstitutiveRelation2012}, the similarity is only superficial, as here the transient constitutive law is local. The non-locality in Eq.~\ref{eq:PP_ODE} is induced by the combination of two processes, the local dilation coupled to the pore flow resisting it.}. For infinitesimal changes of the boundary conditions, i.e for $\delta J \equiv |\Jf - \Ji| \ll 1$, we can work out the ``non-locality" length scale $\hat{\ell}$ over which $\hat{P_\mathrm{p}}$ is affected by the change in $\Jext$. Linearizing Eq.~\ref{eq:PP_ODE} at $\gamma=0^+$ for which $\phi=\phi_i$, we obtain that $\delta \hat{P_\mathrm{p}}= \hat{P_\mathrm{p}}-\hat{P_\mathrm{p}}_\mathrm{i}$ satisfies
\begin{equation}
\partial^2_{\hat{y}} \delta \hat{P_\mathrm{p}} + \frac{\delta \hat{P_\mathrm{p}}}{\hat{\ell}^2} = 0,
\label{eq:PP_ODE_lin}
\end{equation}
with 
\begin{equation}
\hat{\ell} = \sqrt{ \frac{-\gamma_0}{\hat{R}J_\mathrm{i}^2 \phi'_\mathrm{SS}(J_\mathrm{i})}} = \sqrt{ \frac{ \gamma_0}{K\beta \hat{R}(\phi_\mathrm{i})J_\mathrm{i}^{\beta+1}}}.
\label{eq:l}
\end{equation}
The solution of Eq. \ref{eq:PP_ODE_lin}, given the boundary conditions provided in Eq.~\ref{eq:BC}, leads to the particle stress profile
\begin{equation}
	\hat{P_\mathrm{p}}(\hat{y}) = \frac{1}{\Ji} - \left(\frac{1}{\Ji}-\frac{1}{\Jf}\right)\frac{\cosh(\hat{y}/\hat{\ell})}{\hat{\ell}^{-1} \phi_\mathrm{i}^{-1} R(\phi_\mathrm{i})^{-1} \hat{\kappa} \sinh(\hat{H}_\mathrm{i}/\hat{\ell}) + \cosh(\hat{H}_\mathrm{i}/\hat{\ell})},
	\label{eq:initial_stress_profile}
\end{equation}
with $\hat{H}_\mathrm{i}$ the thickness of the granular layer at $\gamma=0^+$. A step change in $J_{\rm ext}$ here leads to an instantaneous change in particle pressure on a depth of order $\hat{\ell}$ below the top wall, while deeper regions remain unchanged relative to the initial steady state at $\Ji$, see Fig. \ref{fig:sketchmodel} (b). In particular, the particle pressure at the bottom of the layer $P_\mathrm{p}(y=0)$ follows:
\begin{equation}
	\frac{P_\mathrm{p}(y=0)}{\Pext} = \frac{\Jf}{\Ji} + \left(\frac{\Jf}{\Ji}-1\right) \frac{1}{\hat{\ell}^{-1} \phi_\mathrm{i}^{-1} R(\phi_\mathrm{i})^{-1} \hat{\kappa} \sinh(\hat{H}_\mathrm{i}/\hat{\ell}) + \cosh(\hat{H}_\mathrm{i}/\hat{\ell})}.
	\label{equ.Pbtr}
\end{equation}
The Darcy-Reynolds model thus predicts that, when $\hat{H}_\mathrm{i}/\hat{\ell}\leqslant1$, a step change in $J_{\rm ext}$ instantly affects the particle pressure all the way to the bottom of the layer. Conversely, in the limit $\hat{H}_\mathrm{i}/\hat{\ell} \to \infty$, the steady-state rheology prediction $P_\mathrm{p}(y=0)/\Pext = \Jf/\Ji $ is recovered. Moreover, we see that $\hat{\ell}$ may diverge at the jamming transition, with an asymptotic behavior $\hat{\ell} \propto \Ji^{-(1+\beta)/2} = (\phi_\mathrm{c} - \phi)^{-(1+\beta)/2\beta}$, if one assumes that $\gamma_0$ remains finite at $\phi_\mathrm{c}$. Investigating the particle pressure profile just after the step in small systems which are close to jamming, i.e. for which $\hat{H}_\mathrm{i}/\hat{\ell} \leqslant 1$, should thus provide an unambiguous way to discriminate the two models. 

Note that in the limit case where $J_\mathrm{i}=0$, i.e. for a granular layer initially at rest, we have $\ell \to \infty$. In such a case, the stress profiles is instantaneously affected over the entire height of the suspension, whatever the system size. We compute the stress profile in this limit case in Appendix B. 


\section{Model testing}
\label{comparison}

In this section, we compare the results of the DEM simulations to the solution of the two models presented above (\S 4). The predictions of the steady-state rheology model are fit-free, since the values of $\beta=0.44$  and $K=0.67$ are set from the steady-state rheological law $\phi_\mathrm{SS}(J)$ obtained from the DEM simulations~\citep{athani2021}. In the Darcy-Reynolds model, we evaluated $\gamma_0$,  with $\gamma_0 = 0.48$ for dilation, and $\gamma_0=0.28$ for compaction separately.

We start by comparing the two models to the macroscopic measurements presented in Fig.~\ref{fig:macro_pheno}.  Overall, the long time evolution of the thickness $H$ of the granular layer, the average particle stress $\bar{P}_\mathrm{p}$ and the depth-averaged shear stress $\bar{\tau}$ (see Fig.~\ref{fig:macro_pheno} b, c and d, respectively) are fairly well predicted by both the steady-state and Darcy-Reynolds models.  While from these results one could conclude that the refinement of the Darcy-Reynolds model over the steady-state rheology model is in practice unnecessary, the value of the particle stress at short times after the step reveals that the prediction of the steady-state rheology model is off by more than one order of magnitude. Conversely, the Darcy-Reynolds model achieves good quantitative predictions even at early strains after the step change of $J_{\rm ext}$, as highlighted in the inset of Fig.~\ref{fig:macro_pheno}. Both models converge and start providing similar results only after a strain scale of order $\gamma_0$.  This result is expected since the strain scale required for stress diffusion to affect the stress profile on a length scale $\hat{\ell}$ is $D(\phi)/\hat{\ell}^2=\gamma_0$.  It is also consistent with the dilatancy law Eq.~\ref{eq:PP_model}, which states that locally, the steady-state rheology is recovered on a strain scale of order $\gamma_0$. These macroscopic observations provide a first illustration of the fundamentally different dynamics for the particle stress in the two models. In the dilation case shown in Fig.~\ref{fig:macro_pheno}, the nonlocal length scale given by the Darcy-Reynolds model is much smaller than the system height ($\hat{H}_\mathrm{i}/\hat{\ell} \approx 0.08$). The change of boundary conditions at the top wall is thus felt instantaneously within the granular layer,  an indication that Darcy-Reynolds coupling is key to capture the early transient dynamics of the granular layer.

To evidence this further, we compare in Fig.~\ref{fig:bottom_p_scatter} the two continuum models to the particle pressure obtained with the DEM simulations at the bottom of the layer and just after the step for many combinations of $\Ji$, $\Jf$, and several system sizes. We have already seen that DEM results can differ from the steady-state rheology prediction at $\gamma=0^{+}$ (thick horizontal grey line), but this difference could be attributed to the finite strain at which DEM results are reported. Importantly, both models and DEM results are now compared at the same strain $\gamma=0.01$. We find that in the steady-state rheology model, stress diffusion from the upper wall can lead to a finite decrease of the particle pressure at $\gamma=0.01$. However, significant discrepancies remain (e.g. purple dashed lines). By contrast, the Darcy-Reynolds model (e.g. purple solid lines) provides surprisingly good quantitative agreement with the DEM simulations over the whole range of parameters. 

In the insets of Fig.~\ref{fig:bottom_p_scatter}, we also show the full pressure profiles at $\gamma=0.01$ for two examples. In the top profile, the step is performed far from jamming (i.e. starting from $J_\mathrm{i}=0.01$). The ``non-locality" length scale $\ell$ is thus much smaller than the system size and as a result, the steady-state and Darcy-Reynolds predictions are undistinguishable at the bottom of the cell. Conversely, in the bottom profile, the step is performed closer to jamming (i.e. starting from $J_\mathrm{i}=0.001$).  This time the ``non-locality" length scale is larger and as a result the bottom particle pressure is instantaneously modified after the step, as predicted by the Darcy-Reynolds model. In both cases however, the Darcy-Reynolds model makes much better predictions for the pressure than the steady-state model near the top wall. Note that these particle stress profiles (along with similar plots at other strain values $\gamma$ and other $\Jf$ values) are used to fit the value of $\gamma_0$, both for compaction ($\Jf=\num{e-4}$), yielding $\gamma_0=0.28$, and dilation ($\Jf=0.1$) yielding $\gamma_0=0.48$. Those values are then kept constant when evaluating the Darcy-Reynolds model predictions.

Our simulation data unambiguously show that the instantaneous response in particle pressure can either be set by the fluid Darcy flow through the particle phase associated with dilation or compaction, or by the steady-state rheology, depending on the value of the parameters involved in the step in $\Jext$. What decides which of these two scenarii dominates? We showed that in the case of the Darcy-Reynolds model, the linearized pressure ODE (\ref{eq:PP_ODE_lin}) predicts that the Darcy flow is associated with a length scale $\hat{\ell}$ corresponding to the typical length over which a dilation or compaction (and the associated Darcy flow) is instantaneously initiated when the boundary conditions on the top wall are modified. A natural expectation is thus that, even in the non-linear regime, the pressure response is dominated by the Darcy flow when the initial height of the system $\hat{H}_\mathrm{i}$ is of order $\hat{\ell}$. On the contrary, when $\hat{H}_\mathrm{i}/\hat{\ell}\gg 1$, we may expect that the steady rheology sets the pressure level in most of the system except a thin layer of height $\hat{\ell}$ below the top wall.

\begin{figure}
\centering
\includegraphics[width=12cm]{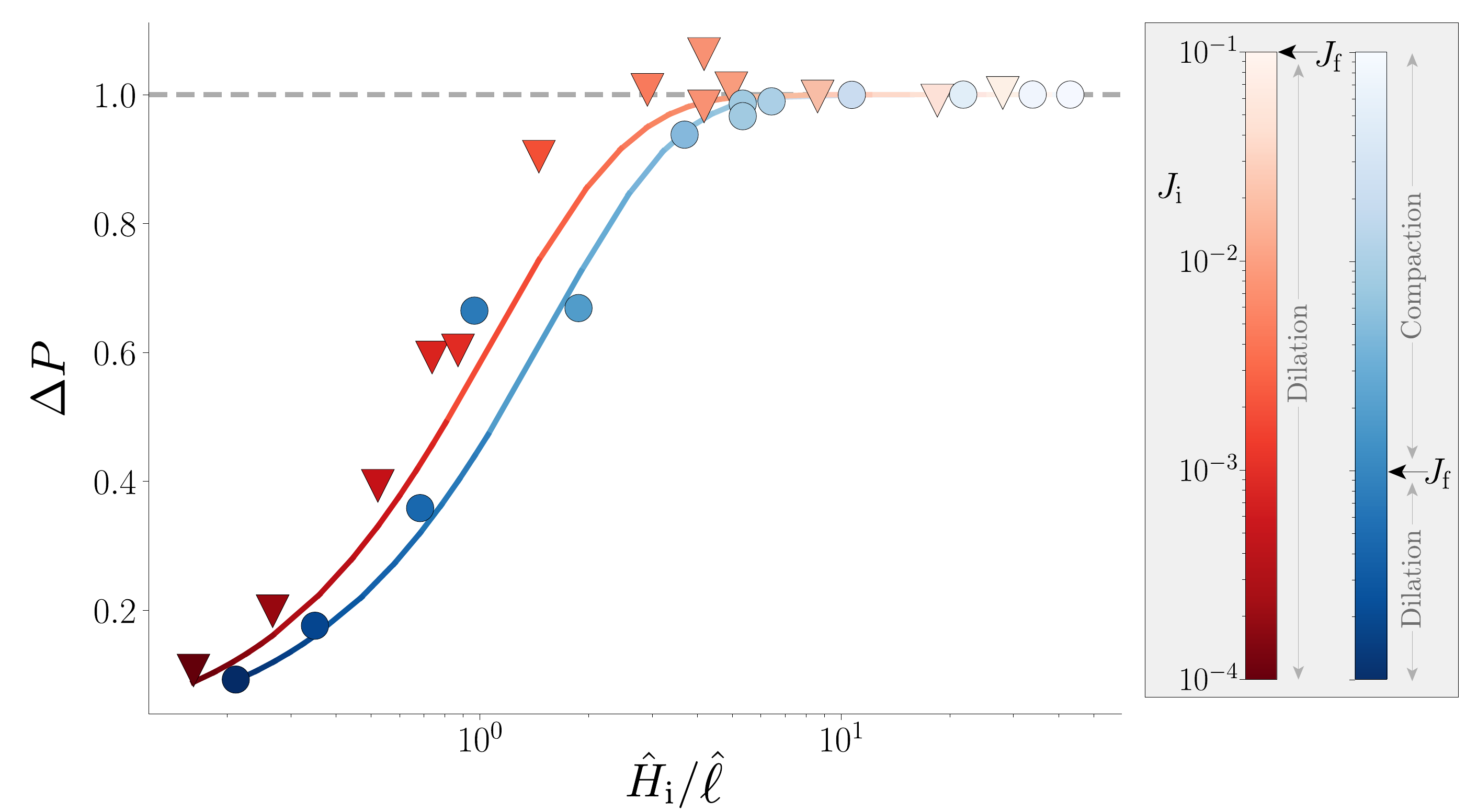}
\caption{The rescaled bottom particle pressure $\Delta P= \left(P_\mathrm{p}(y=0)/P_{\rm ext} - 1\right)/\left(1-\Ji/\Jf\right) + 1$ plotted versus $\hat{H}_\mathrm{i}/\hat{\ell}$ at $\gamma = 0.01$ shows that the boundary between the steady-state rheology and the Darcy-Reynolds dominated regime is set by the nonlocal length scale $\hat{\ell}$, and that the steady-state rheology only applies when $\hat{H}_\mathrm{i}/\hat{\ell} \gg 1$.  Symbols: DEM simulations, Solid lines: Darcy-Reynolds model, Grey dashed line: steady-state rheology model at $\gamma=0^{+}$.}  
\label{fig:bottom_p_vs_H}       
\end{figure}

In Fig.~\ref{fig:bottom_p_vs_H} we therefore show as a function of $\hat{H}_\mathrm{i}/\hat{\ell}$ the rescaled bottom pressure 
$\Delta P = \left(P_\mathrm{p}(y=0)/P_{\rm ext} - 1\right)/\left(1-\Ji/\Jf\right) + 1$. This rescaling is useful as for any step in $\Jext$, $\Delta P$ is bounded such that $0 < \Delta P< 1$, with $\Delta P= 1$ corresponding to the prediction of the steady rheology at $\gamma=0^{+}$. We find that for all the combination of $J_{\rm i}$ and $J_{\rm f}$ investigated, data collapse on a master curve, where $\Delta P$  monotonically increases when plotted versus $\hat{H}_\mathrm{i}/\hat{\ell}$, and saturates at $1$ for $\hat{H}_\mathrm{i}/\hat{\ell}\gtrsim 10$. This confirms that the boundary between the steady-rheology dominated and Darcy-Reynolds dominated regimes is set by $\hat{\ell}$. In Fig.~\ref{fig:bottom_p_vs_H}, we also show the prediction of the Darcy-Reynolds model (solid lines). The agreement with DEM simulation data is good, over the whole range of parameter investigated, that is for $\hat{H}_\mathrm{i}/\hat{\ell}$ varying over three orders of magnitude and $\Ji$ varying over four orders of magnitude. The fact that the best fits of the particle stress profiles are obtained for two distinct values of $\gamma_0$ in dilation ($\gamma_0=0.48$) and compaction ($\gamma_0=0.28$) implies a $\Jf$-dependence of $\hat{\ell}$ which is ignored in the linearized problem, but probably cannot be in the cases considered here, for which $\Ji-\Jf$ is of order of or even larger than $\Ji$. 

\section{Discussion \& Conclusion}
\label{conclu}

Through DEM simulations, we investigated the transient rheological behavior of a neutrally buoyant suspension under pressure imposed conditions, subject to a sudden change in shear rate or external pressure. We compare these simulations with two competing continuum two-phase models: the standard Suspension Balance Model (SBM) which assumes the steady-state rheology to be valid at all times, and a ``Darcy-Reynolds" model in which the volume fraction locally relaxes towards its steady-state value on a strain scale $\gamma_0$, by analogy with Reynolds dilatancy in soil mechanics. This study shows that the early stress response of the suspension is not set by the steady-state rheological flow rules, but instead arise from the Darcy back-flow resulting from the geometrically imposed dilation rate of the granular phase (\ref{eq:phi_SS}). 
Before discussing the consequences of these results, let us recall that they were obtained for a dynamics which is essentially one-dimensional.
This is because we considered a model setup where dilation occurs in only one direction and shear is homogeneous in the sample.
Some results we obtain may be quantitatively modified in a less idealized setup, as for instance a non-uniform shear rate could locally modulate the Reynolds dilatancy.
Nonetheless we believe that the qualitative picture which emerges from our results should remain.
Indeed, our results have several important consequences and implications: 

First, our study extends the domain of application of the Reynolds-like dilatancy law (\ref{eq:PP_model})  proposed by \citet{pailha_two-phase_2009}, which was introduced to describe the transient dilation/compaction of avalanches of an initially dense/loose sediment under gravity. Here, we provide evidence that this law also applies  for continuously sheared suspensions below $\phi_\mathrm{c}$, when the flow parameters (external pressure, shear rate) are suddenly changed. The transient migration and stresses are quantitatively captured over a wide range of control parameter, $J\in[10^{-4}-10^{-1}]$, and system sizes, $H/a \in [33-140]$. Our study thereby shows that the concepts of Reynolds dilatancy and shear-induced migration can be described within a unique framework provided by the Darcy-Reynolds model (\ref{eq:PP_ODE}). 

Second, our study reveals that after a sudden change of flow parameters, the stress levels inside the suspension can be very different than that predicted by the steady-state rheology. More precisely, the Darcy-Reynolds model gives rise to a nonlocal length scale $\ell$, which scales with the particle size and diverges algebraically at jamming (\ref{eq:l}). This length scale corresponds to the distance from the free boundary over which a finite rate of dilation/compaction occurs after a step change of flow conditions. In this region, the stress level is thus fixed, not by the steady-state rheology, but by the Darcy fluid pressure gradient resulting from this dilation/compaction rate.

\begin{figure}
\centering
  \includegraphics[width=9 cm]{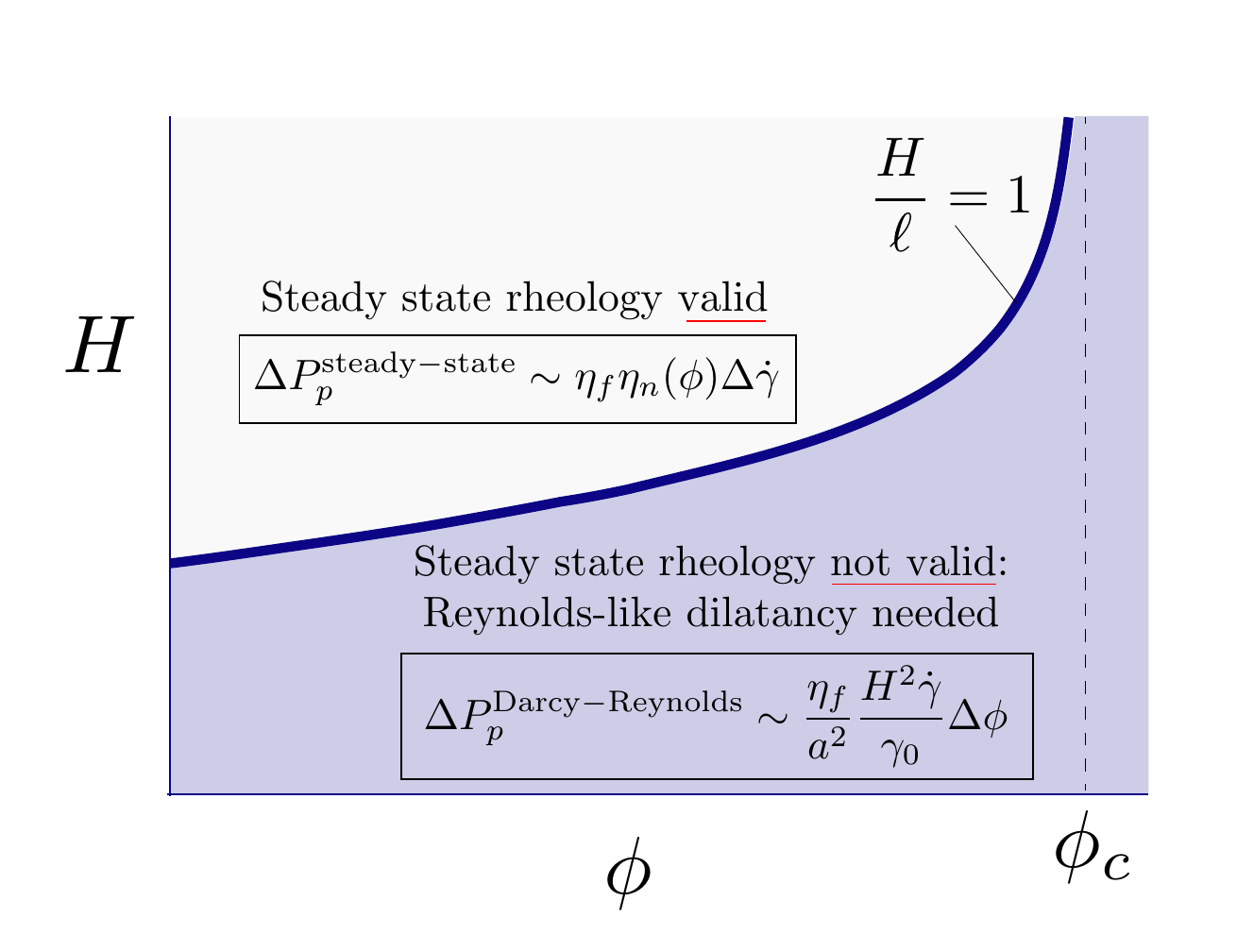}
\caption{Domain of application of the steady-state rheology model and adequate scaling for the particle stress for the early response of a suspension of initial  volume fraction $\phi$ and typical size $H$, subject to a sudden change of flow condition $\Delta J$. Here $\Delta \dot{\gamma}$ is the shear rate jump and $\Delta \phi= \vert \phi-\phi_\mathrm{SS}(J_\mathrm{f}) \vert$.   }
\label{fig:flow_regimes}       
\end{figure}

Accounting for this mechanism is key to predict the stress levels at early strains ($\gamma \le \gamma_0$), 
which we quantify by the difference between the external imposed pressure and the particle pressure in the bulk, $\Delta P_\mathrm{p}$. 
As summarized in the flow regime diagram sketched in Fig.~\ref{fig:flow_regimes}, for a suspension of volume fraction $\phi$, when the system size $H$ is much larger than $\ell$, 
$\Delta P_\mathrm{p}$ is well predicted by the steady-state rheology of the suspension.
In a situation where the shear rate is suddenly modified from $\dot\gamma$ to $\dot\gamma + \Delta \dot\gamma$, the quadi-Newtonian steady-state rheology predicts
$\Delta P_\mathrm{p}^{\rm{steady-state}}\sim \eta_\mathrm{f} \eta_\mathrm{n}(\phi) \Delta \dot{\gamma}$. 
On the contrary, when $H \le \ell$, the transient pressure is set by the drag of the fluid on the particle phase generated by the Darcy flow coming from dilation or compaction, 
which spans the whole system. 
It therefore depends on the particle and system sizes. 
As the pressure gradient across the system scales as $\Delta P_\mathrm{p}/H$, from Eq. \ref{eq:PP_ODE} we can evaluate the Darcy-Reynolds scaling 
$\Delta P_\mathrm{p}^{\rm{Darcy-Reynolds}}\sim \eta_\mathrm{f} \dot{\gamma} H^2 \Delta \phi/ a^2 \gamma_0$, where $\Delta \phi= \vert \phi-\phi_\mathrm{SS}(J_\mathrm{f}) \vert$ is the amount of dilation/compaction between the initial and final states. 
The boundary between the steady-rheology dominated and Darcy-Reynolds dominated regimes is thus set by the length scale ratio $H/\ell=\mathcal{O}(1)$.  This boundary also corresponds to the conditions for which the steady-state and Darcy-Reynolds prediction for the stress are equal, i.e. $\Delta P_\mathrm{p}^{\rm{Darcy-Reynolds}}/\Delta P_\mathrm{p}^{\rm{steady-state}}=\mathcal{O}(1)$.  Indeed, for a small flow perturbation $\Delta \dot{\gamma}\approx \dot{\gamma} \Delta J /J^2$ while, $\Delta \phi \approx K \beta J^{\beta-1}\Delta J$, which gives $\Delta P_\mathrm{p}^{\rm{Darcy-Reynolds}}/\Delta P_\mathrm{p}^{\rm{steady-state}}=H/\ell$.

Our demonstration of the validity of the Reynolds-like dilatancy law (\ref{eq:PP_model}) for flow below $\phi_\mathrm{c}$ and the identification of the nonlocal length scale $\ell$ for estimating the particle stress should help to better  understand the transient dynamics of suspensions, as observed during impacts~\citep{nicolas2005,peters2013,grishaev2015,schaarsberg2016, boyer2016}, submarine avalanches~\citep{rondon2011granular,topin2012collapse,iverson2012elementary,bougouin2018granular,montella2021two}, or unsteady two-phase flows in general~\citep{kulkarni2010particle,snookDynamicsShearinducedMigration2016,saint2019x,d2021viscous}. For instance, the Darcy-Reynolds model was successfully used to described the impact of a sphere on a suspension initially prepared above $\phi_\mathrm{c}$~\citep{jerome2016unifying}. However, the behavior of the system below $\phi_\mathrm{c}$ could not be described by the quasi-static approach used in this study. Similarly, when prepared just below $\phi_\mathrm{c}$, suspension drops impacting a rigid plane observe large spreadings that cannot be captured using the steady-state viscosity of the suspension \citep{jorgensen2020}. For such small systems close to jamming, it is likely that $H/\ell<1$, for which the  Darcy-Reynolds scaling applies. Note that these configurations are apparently under volume imposed conditions. However, the presence of a free surface allows a slight dilation of the granular network after the impact. We thus anticipate that in these configurations too, the level of stress right after the impact should originate, not from the steady-state rheology, but from the transient geometrical dilation of the granular phase and the associated pore pressure feedback effect. 

Our study could also have implication  to rationalize the transient dynamics of shear thickening suspensions. Shear thickening suspensions are characterized by a stress-dependent critical packing fraction~\citep{seto2013,wyart2014}. Upon a sudden change of boundary conditions, an initially unjammed suspension may be driven above its maximum packing fraction and shear-jam. This shear-jammed regime exhibits fascinating transient features such as impact-activated solidification and traveling  ``jamming front'' \citep{waitukaitis2012impact,han2016high,han2018shear}, that cannot be described using steady-state rheological flow rules. Predicting the resistive stress in this shear-jammed regime is an important issue which is still largely unresolved. Most existing models assume a minimal description in which the medium jams in a finite strain \citep{waitukaitis2012impact,han2016high,han2018shear} without considering two-phase flow coupling (see however \citep{jerome2016unifying} and \citep{brassard2020viscous}). However, in such configurations standing above $\phi_\mathrm{c}$ where the nonlocal length scale $\ell$ diverges, one expects the Darcy-Reynolds coupling to play a major role.

Finally, an important open question concerns the microscopic origin of the Reynolds-like dilatancy law (\ref{eq:PP_model}) proposed by~\cite{pailha_two-phase_2009}. In the quasistatic regime, Reynolds dilatancy is usually explained from geometric/kinematic arguments through the introduction of a dilatancy angle~\citep{reynolds1885lvii,wood1990soil}. More recently, this law was reinterpreted as a normal stress relaxation law in a medium of finite compressibility~\citep{bouchut2016two,lee2021two,montella2021two}, in line with the ``Reynolds pressure'' concept of~\cite{ren2013reynolds}. Compared to the steady-state rheology model, the Reynolds-like dilatancy law requires the microstructure to reorganize in a strain scale $\gamma_0$ of order $1$. Capturing the microstructure dynamics in a linear relaxation of a scalar variable is certainly an over simplification, as one probably needs to generally consider a tensorial descriptor of the microstructure~\citep{chackoShearReversalDense2018,gillissenModelingSphereSuspension2018}.
However, in the case of a simple shear in a direction constant in time,  a scalar relaxation has already proven useful to describe transient responses~\citep{mari_nonmonotonic_2015,chacko_dynamic_2018,hanShearFrontsShearthickening2018}.
It is worth noting that in \cite{pailha_two-phase_2009}, for dilation experiments starting from above the jamming point ($\phi>\phi_\mathrm{c}$), Reynolds dilatancy follows $\gamma_0 \approx 0.3$. This value is intriguingly close to the values of $\gamma_0$ we find for our $\phi<\phi_\mathrm{c}$ conditions. It is thus tempting to conjecture that the value for $\gamma_0$ is independent of $\phi$, or at least only weakly dependent on $\phi$, around the jamming transition. This would mean that the classical Reynolds dilatancy of the critical state theory is just a limiting case of a more general dilatancy law valid over a wide range of volume fractions.

\section*{Acknowledgements}
This work was funded by ANR ScienceFriction (ANR-18-CE30-0024).

\section*{Declaration of interests} 
The authors report no conflict of interest.

\appendix

\section{Numerical method}
\label{app:numerical_method}

As described in Sec.~\ref{sec:numerics}, the granular suspension considered in this study is neutrally buoyant and we work in vanishing Stokes number and Reynolds number regime; consequently, the equation of motion is just the force balance between hydrodynamic forces and contact forces for bulk particles, and between hydrodynamic forces, contact forces and the externally applied force for wall particles.
Contact forces are modelled using a system of springs and dashpots~\citep{cundallDiscreteNumericalModel1979}. 
Using the notation $\bm{F}_\mathrm{C} \equiv \left(\bm{f}_{\mathrm{C},1}, \dots, \bm{f}_{\mathrm{C},N}, \bm{t}_{\mathrm{C},1}, \dots, \bm{t}_{\mathrm{C},N}\right)$ for the $N d(d+1)/2$ vector of contact forces and torques on each particle (in spatial dimension $d$), we have
\begin{equation}
\bm{F}_\mathrm{C}
=
- \bm{R}^\mathrm{C}_\mathrm{FU} \cdot \bm{U}
+ 
\bm{F}_\mathrm{C,S}
\end{equation}
with $\bm{U}$ (resp. $\bm{\Omega}$) the vector of velocities (resp. angular velocities) for each particle, $\bm{R}^\mathrm{C}_\mathrm{FU}$ the resistance matrix associated to dashpots, and $\bm{F}_\mathrm{C,S}$ the part of contact forces coming from springs.
For hydrodynamic forces and torques $\bm{F}_\mathrm{H}$, we have~\citep{jeffreyCalculationLowReynolds1992}
\begin{equation} \label{eqhydro}
\bm{F}_\mathrm{H}
=
- \bm{R}^\mathrm{H}_\mathrm{FU} \cdot \bm{U}^\prime
+ 
\bm{R}_\mathrm{FE}:\bm{E}^\infty
\end{equation}
with $\bm{U}^\prime = \bm{U} - \bm{U}^\infty$ the vector of non-affine velocities and angular velocities for each particle, where $\bm{U}^\infty$ are the background velocities and angular velocities evaluated at the particles centers, 
and $\bm{E}^\infty$ the symmetric part of the imposed velocity gradient $\nabla \bm{U}^\infty$.
The resistance matrices $\bm{R}^\mathrm{H}_\mathrm{FU}$ and $\bm{R}_\mathrm{FE}$ contain Stokes drag and regularized lubrication forces at leading order in particle separation~\cite{mari2014shear}.

The equation of motion is thus
\begin{equation}\label{eq:eq_motion_full}
    \bm{F}_\mathrm{H} + \bm{F}_\mathrm{C} + \bm{F}_\mathrm{Ext} = 0\, ,
\end{equation}
with $\bm{F}_\mathrm{Ext}$ the external force/torque applied on each particle. Of course, $\bm{F}_\mathrm{Ext}$ takes non-zero values only for the wall particles.

As the simulated system (see Fig. \ref{system2}) comprises of frozen wall and bulk particles suspended in the fluid, we separate the total resistance matrix as
\begin{equation} \label{eqresmatrix}
\bm{R}^\mathrm{C}_\mathrm{FU} + \bm{R}^\mathrm{H}_\mathrm{FU} \equiv \bm{R}_\mathrm{FU}
=
\begin{pmatrix}
\bm{R}_\mathrm{FU}^\mathrm{bb} & \bm{R}_\mathrm{FU}^\mathrm{bw}\\
\bm{R}_\mathrm{FU}^\mathrm{wb} & \bm{R}_\mathrm{FU}^\mathrm{ww}
\end{pmatrix}
\end{equation}
The matrices $\bm{R}_\mathrm{FU}^\mathrm{bb}$, $\bm{R}_\mathrm{FU}^\mathrm{bw}$, $\bm{R}_\mathrm{FU}^\mathrm{wb}$ and $\bm{R}_\mathrm{FU}^\mathrm{ww}$ indicate the hydrodynamic resistance matrices including bulk-bulk, bulk-wall, wall-bulk and wall-wall interactions respectively.
Similarly, the non-affine velocities have also been separated into bulk particle velocities $\bm{U}^\mathrm{b} - \bm{U}^{\infty}$ and wall particle velocities $\bm{U}^\mathrm{w} - \bm{U}^{\infty}$,
\begin{equation} \label{equdecom}
\bm{U}^\prime 
=
\begin{pmatrix}
\bm{U}^\mathrm{\prime b}\\
\bm{U}^\mathrm{\prime w}
\end{pmatrix}\, .
\end{equation}
We can then solve the equations of motion, Eq.~\ref{eq:eq_motion_full} for the bulk velocities, to get
\begin{equation}\label{eq:bulk_velocities}
    \bm{U}^\mathrm{\prime b} = {\bm{R}_\mathrm{FU}^{\mathrm{bb}}}^{-1} \cdot \left[\bm{K}^\mathrm{b} - \bm{R}_\mathrm{FU}^\mathrm{bw}\cdot \bm{U}^\mathrm{\prime w}\right]
\end{equation}
with
\begin{equation}
    \begin{pmatrix}
    \bm{K}^\mathrm{b}\\
    \bm{K}^\mathrm{w}
    \end{pmatrix} = \bm{R}_\mathrm{FE}:\bm{E}^\infty + \bm{F}_\mathrm{C,S} - \bm{R}^\mathrm{C}_\mathrm{FU} \cdot \bm{U}^\infty
\end{equation}
We further decompose the wall non-affine velocity in horizontal and vertical components $\bm{U}^\mathrm{\prime w} = \bm{U}^\mathrm{\prime w}_\mathrm{h} + v_y \bm{Y}^\mathrm{w}$, where $\bm{Y}^\mathrm{w}$ is the vector corresponding to a unit non-affine vertical velocity  for particles belonging to the upper wall, and vanishing non-affine velocity for particles of the lower wall.
Thus the scalar $v_y = \partial_t H$ is the upper wall vertical speed.

Injecting Eq.~\ref{eq:bulk_velocities} in the wall part of Eq.~\ref{eq:eq_motion_full}, and using Eq.~\ref{equdecom}, we get
\begin{equation}
   v_y \bm{B}\cdot \bm{Y}^\mathrm{w} = \bm{K}^\mathrm{w} - \bm{R}_\mathrm{FU}^\mathrm{wb}\cdot {\bm{R}_\mathrm{FU}^{\mathrm{bb}}}^{-1}\cdot \bm{K}^\mathrm{b} - \bm{B}\cdot \bm{U}^\mathrm{\prime w}_\mathrm{h} + \bm{F}_\mathrm{Ext}\, .
\end{equation}
Taking the dot product with $\bm{Y}^\mathrm{w}$, and using $\bm{Y}^\mathrm{w}\cdot \bm{F}_\mathrm{Ext} = \Pext l_x$ we get the vertical wall velocity
\begin{equation}
     v_y = \frac{\bm{Y}^\mathrm{w} \cdot\left[ \bm{K}^\mathrm{w} - \bm{R}_\mathrm{FU}^\mathrm{wb}\cdot {\bm{R}_\mathrm{FU}^{\mathrm{bb}}}^{-1}\cdot \bm{K}^\mathrm{b} - \bm{B}\cdot \bm{U}^\mathrm{\prime w}_\mathrm{h}\right] + \Pext l_x}{\bm{Y}^\mathrm{w}\cdot \bm{B}\cdot \bm{Y}^\mathrm{w}}\, .
\end{equation}
Imposing that the vertical component of the upper wall velocity is $v_y$ ensures that the total force on the upper wall vanishes.


\section{\label{app:solution_phic}Stress profile from uniform state}

\subsection{Jammed initial state}
As noted in Sec.~\ref{sec:modelling}, within the Darcy-Reynolds model upon change of $\Jext$ at $\gamma=0$, the viscous number is instantaneously changing in the part of the suspension just below the top wall, 
on a length scale $\hat{\ell}$. When the initial state is at $\Ji=0$, $\hat{\ell}$ however diverges, and the response is qualitatively different from the one exposed in Eq.~\ref{eq:initial_stress_profile}.
We can illustrate this as Eq.~\ref{eq:PP_ODE} has an essentially analytically tractable solution for an initial profile $\phi_\mathrm{i} = \phi_\mathrm{c}$ when $\beta=1/2$.

From Eq.~\ref{eq:PP_ODE}, first with an initially arbitrary $\beta$, for $\phi_\mathrm{i} = \phi_\mathrm{c}$ with Eq.~\ref{eq:phi_SS} the stress satisfies
\begin{equation}
	\partial^2_{\hat{y}}\hat{P}_\mathrm{p} + \gamma_0^{-1} \hat{R}(\phi_\mathrm{c}) K {\hat{P}_\mathrm{p}}^{-\beta} = 0.
\end{equation}
Using $\partial_{\hat{y}}\hat{P}_\mathrm{p}$ as an integrating factor, we get
\begin{equation}
	\frac{1}{2}\left[\partial_{\hat{y}}\hat{P}_\mathrm{p} \right]^2 = - \frac{\hat{R}(\phi_\mathrm{c}) K}{\gamma_0(1-\beta)} {\hat{P}_\mathrm{p}}^{-\beta+1} + \frac{C}{2},
\end{equation}
with $C$ an integration constant. Remembering that from a jammed configuration we expect dilation, that is $\partial_{\hat{y}}\hat{P}_\mathrm{p} < 0$, leads to
\begin{equation}
	\partial_{\hat{y}}\hat{P}_\mathrm{p} = -\sqrt{ - 2 \frac{\hat{R}(\phi_\mathrm{c}) K}{\gamma_0(1-\beta)} {\hat{P}_\mathrm{p}}^{-\beta+1} + C}.
	\label{eq:first_order_intermediate}
\end{equation}
To go further we assume $\beta=1/2$, close to the actual measured value, and separating variables we can integrate to get
\begin{equation}
	\hat{y} = \frac{\sqrt{2} \gamma_0^2}{3\hat{R}(\phi_\mathrm{c})^2 K^2} \left[ \frac{\hat{R}(\phi_\mathrm{c}) K}{\gamma_0} \sqrt{\hat{P}_\mathrm{p}} + C \right]
				\sqrt{ - 2 \frac{\hat{R}(\phi_\mathrm{c}) K}{\gamma_0} \sqrt{\hat{P}_\mathrm{p}} + C} + C',
\end{equation}
with a new constant of integration $C'$. Because of the no-flux boundary condition at the bottom wall, $\partial_{\hat{y}}\hat{P}_\mathrm{p}(0) = 0$, which with Eq.~\ref{eq:first_order_intermediate} 
imposes that $\sqrt{ - 2 \frac{\hat{R}(\phi_\mathrm{c}) K}{\gamma_0} \sqrt{\hat{P}_\mathrm{p}} + C} = 0$ for $\hat{y}=0$. We therefore have $C'=0$, as well as $C = 2\hat{R}(\phi_\mathrm{c})K\sqrt{\hat{P}_\mathrm{p}(0)}/\gamma_0 >0$.
Introducing $r=\sqrt{\hat{P}_\mathrm{p}}$ and $A=\hat{R}(\phi_\mathrm{c}) K/\gamma_0$, we then get a cubic equation for $r$,
\begin{equation}
 0 = \frac{2A}{3}r^3 + Cr^2 - \frac{C^3}{3A^2} + \frac{3\hat{y}^2}{2} \equiv Q(r).
\end{equation}
The determinant of this polynomial is
\begin{equation}
\Delta = -\frac{16A^2}{3}\hat{y}^2\left(-\frac{2C^3}{9A^2} + \hat{y}^2\right).
\end{equation}
For small enough $\hat{y}$, $\Delta > 0$ (and $\Delta=0$ for $\hat{y}=0$), which corresponds to having three real solutions. 
Because for $\hat{y}=0$, $Q(0)<0$ and its cubic and quadratic terms have positive coefficients, we have two negative solutions and a positive one. 
As $r$ must be positive, the latter is the correct solution. For $\hat{y}=\sqrt{2}C^{3/2}/(3A)$, the determinant vanishes, and this solution vanishes. 
Because we want the solution to be positive up to the top wall, this gives us a lower bound for $C$, $C > (3A\hat{H}/\sqrt{2})^{2/3}$.

The three roots of $Q$ are, with $k=0,1,2$
\begin{equation}
r_k  = - \frac{C}{2A} \left[ 1 + 2 \cos \left\{ \frac{1}{3} \left[ \varphi \left(\frac{3A}{C^{3/2}} \hat{y} \right) + (2k+1)\pi \right] \right\} \right],
\end{equation}
with $\varphi$ such that $e^{i \varphi(x)} = 1 - x^2 - i x\sqrt{2-x^2}$.  
We can identify the correct root with the case $\hat{y} = 0$, for which we know that there are two degenerate negative solutions and the positive one we are interested in. 
Indeed, as $\varphi(0) = 0$, we have that $r_0 = r_2 = -\frac{C}{A}$ for $\hat{y} = 0$, while $r_1 = \frac{C}{2A}$, which implies that $r_1$ is the root we are looking for.
As a result, replacing $A$ by its value in terms of the problem parameters, we have the stress profile at $\gamma=0$
\begin{equation}
\hat{P}_\mathrm{p}(\hat{y}) =  \frac{\gamma_0^2 C^2}{4\hat{R}(\phi_\mathrm{c})^2 K^2} \left\{ 1 - 2 \cos \left[ \frac{1}{3} \varphi \left(\frac{3\hat{R}(\phi_\mathrm{c}) K}{\gamma_0 C^{3/2}} \hat{y} \right) \right] \right\}^2.
\label{eq:stress_profile_phic}
\end{equation}
We can then pick $C$ to satisfy the top wall boundary condition defined in Eq.~\ref{eq:BC}, but this must be done numerically as there is no mathematically 
closed expression for $C$. A few important things can however be inferred from Eq.~\ref{eq:stress_profile_phic}. 

First, although a cosine appears in the r.h.s., it of course does not mean that the stress profile is non-monotonic. 
Indeed, as noted earlier $C$ is bounded from below by $(3A\hat{H}/\sqrt{2})^{2/3}$, which means that the argument of $\varphi$ in Eq.~\ref{eq:stress_profile_phic} 
is satisfying $\frac{3\hat{R}(\phi_\mathrm{c}) K}{\gamma_0 C^{3/2}} \hat{y} < \sqrt{2}$. 
It is easy to show that $\varphi(0) = 0$ and $\varphi(\sqrt{2}) = \pi$, so that the cosine in r.h.s of Eq.~\ref{eq:stress_profile_phic} is decreasing from the bottom wall to the top wall, 
its argument spanning at most the interval $[0, \pi/3]$.
Moreover, the stress at the top of the cell must be different from the stress at the bottom, and from the argument of the cosine that it implies that $C$ must scale as $\hat{H}^{2/3}$ for large $\hat{H}$.

\begin{figure}
\centering
  \includegraphics[width=0.6\columnwidth]{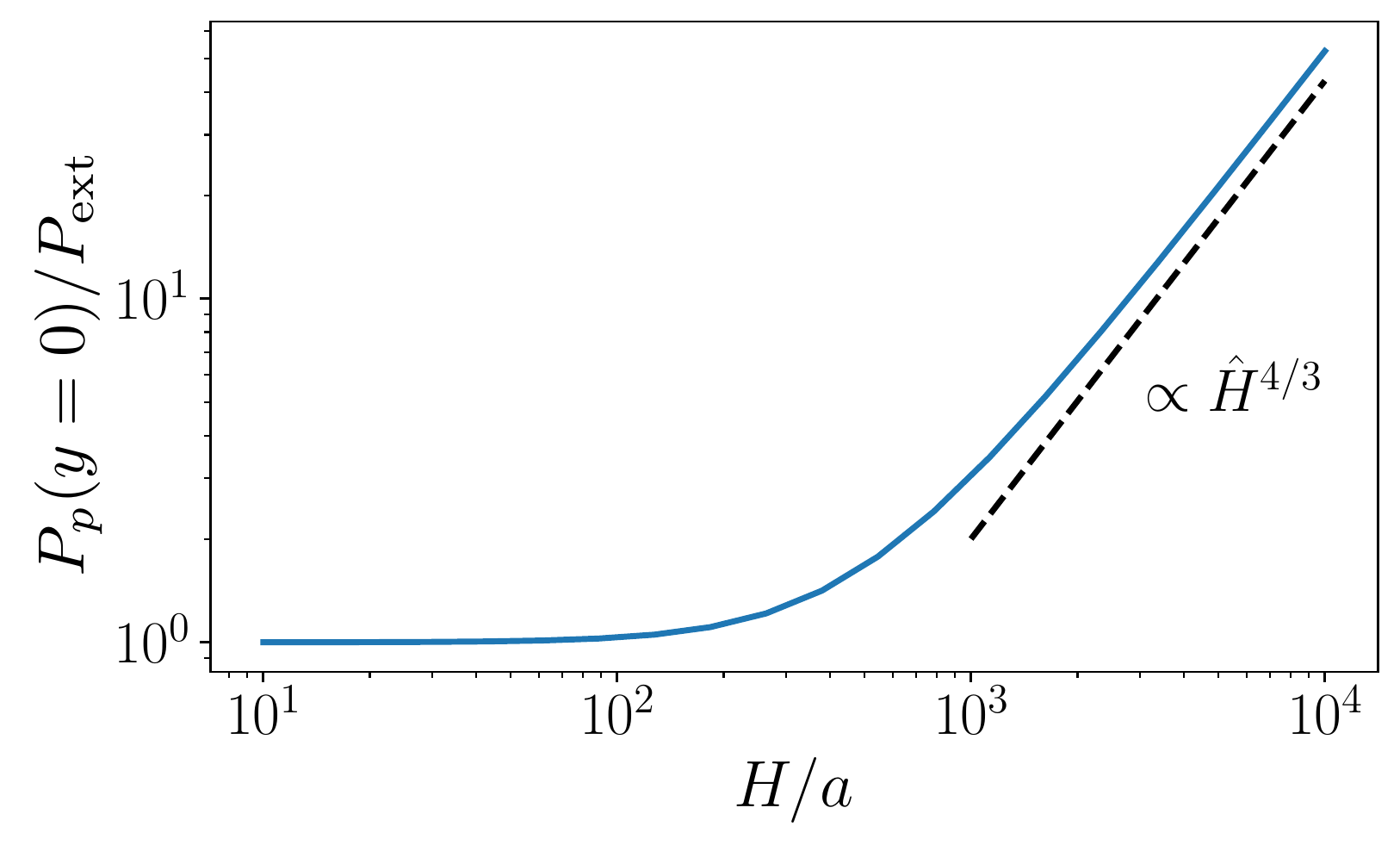}
\caption{Particle pressure at the bottom wall $\hat{P}_\mathrm{p}(y=0)$ at $\gamma = 0^+$ scaled by the imposed pressure at the top wall $P_{\rm ext}$, as a function of the system height $\hat{H} = H/a$, when the initial configuration is jammed ($\Ji=0$), from Eq.~\ref{eq:stress_profile_phic} . The pressure scales as $\hat{H}^{4/3}$ in the large $\hat{H}$ limit.}
\label{fig:pressure_scaling}       
\end{figure}

Second, we expect from the divergence of $\hat{\ell}$ when $\phi_\mathrm{i}=\phi_\mathrm{c}$ that the stress profile is instantaneously modified at all points in the system at $\gamma=0$. 
The fact that a branch of cosine taking $\hat{y}$ in its argument appears in Eq.~\ref{eq:stress_profile_phic} is pointing towards such a scenario, but because of the nonlinear 
behavior of $\varphi$ it is not immediately apparent that the profile is not flat in the bottom of the cell. We know that $\partial_{\hat{y}}\hat{P}_\mathrm{p}(0)=0$ by construction 
(which can easily be verified on Eq.~\ref{eq:stress_profile_phic}), so we have to evaluate the second derivative at the bottom wall to be informed about the flatness of the profile there. 
A quick calculation, using that $\varphi'(0) = -\sqrt{2}$, shows that the curvature of the stress profile at the bottom wall is
\begin{equation}
\partial^2_{\hat{y}}\hat{P}_\mathrm{p}(0)=- \frac{2}{C} \propto \hat{H}^{-2/3},
\end{equation}
which is to be compared with the $\propto \exp(-\hat{H}/\hat{\ell})$ for $\phi_\mathrm{i} < \phi_\mathrm{c}$.
This weak scaling implies that the stress at the bottom will be height dependent, which is easy to verify, as
\begin{equation}
	\hat{P}_\mathrm{p}(0) =  \frac{\gamma_0^2 C^2}{4\hat{R}(\phi_\mathrm{c})^2 K^2} \propto \hat{H}^{4/3},
\end{equation}
here again in stark contrast to the case $\Ji>0$, for which the stress at the bottom is initially converging to $-1/\Ji$ exponentially in $\hat{H}/\hat{\ell}$.
This approach to the asymptotic behavior is shown in Fig.~\ref{fig:pressure_scaling}.

\subsection{Flowing initial state}

If the initial state is a steady state under viscous number $\Ji$, we again can derive the stress profile at $\gamma=0$ for a rheology given by Eq.~\ref{eq:phi_SS} in the case $\beta=1/2$.
Injecting Eq.~\ref{eq:phi_SS} in Eq.~\ref{eq:PP_ODE}, with a uniform solid fraction $\phi_\mathrm{i} = \phi_\mathrm{SS}(\Ji)$, we get
\begin{equation}
	\partial^2_{\hat{y}}\hat{P}_\mathrm{p} + \gamma_0^{-1} \hat{R}(\phi_\mathrm{i}) K \left[-\sqrt{\Ji} + \frac{1}{\sqrt{\hat{P}_\mathrm{p}}}\right] = 0.
\end{equation}
Using $\partial_{\hat{y}}\hat{P}_\mathrm{p}$ as an integrating factor, we get
\begin{equation}
	\frac{1}{2}\left[\partial_{\hat{y}}\hat{P}_\mathrm{p} \right]^2 = \gamma_0^{-1} \hat{R}(\phi_\mathrm{i}) K \left[2\left(
	\sqrt{\hat{P}_\mathrm{p}(0)} - \sqrt{\hat{P}_\mathrm{p}}\right) + \sqrt{\Ji}\left(\hat{P}_\mathrm{p} - \hat{P}_\mathrm{p}(0)\right) \right],
\end{equation}
where we used the bottom wall boundary condition $\partial_{\hat{y}}\hat{P}_\mathrm{p}(0) = 0$ and we recall that $\hat{P}_\mathrm{p}(0)$ is the particle pressure at the bottom wall. This leads to
\begin{equation}
	\partial_{\hat{y}}\hat{P}_\mathrm{p} = \pm \sqrt{ 2 \gamma_0^{-1} \hat{R}(\phi_\mathrm{i}) K \left[2\left(\sqrt{\hat{P}_\mathrm{p}(0)} - \sqrt{\hat{P}_\mathrm{p}}\right) + \sqrt{\Ji}\left(\hat{P}_\mathrm{p} - \hat{P}_\mathrm{p}(0)\right)\right]}.
	\label{eq:stress_ode_flat_profile}
\end{equation}
Here the choice of the r.h.s. sign depends on the change of $\Jext$. 
A dilation, with $\Jf > \Ji$, implies $\partial_{\hat{y}}\hat{P}_\mathrm{p} \leq 0$, whereas a compaction, with $\Jf < \Ji$, implies $\partial_{\hat{y}}\hat{P}_\mathrm{p} \geq 0$.
Introducing 
\begin{equation}
    g(\hat{P}_\mathrm{p}) =\sqrt{2\left(\sqrt{\hat{P}_\mathrm{p}(0)} - \sqrt{\hat{P}_\mathrm{p}}\right) + \sqrt{\Ji}\left(\hat{P}_\mathrm{p} - \hat{P}_\mathrm{p}(0)\right)}\, ,
\end{equation} 
we can integrate Eq.~\ref{eq:stress_ode_flat_profile} by separation of variables, yielding
\begin{equation}
\hat{y} =  \sqrt{\frac{\gamma_0}{2\hat{R}(\phi_\mathrm{i})K\Ji}} \left\{g(\hat{P}_\mathrm{p}) + \Ji^{-1/4}
	\ln\left[\frac{ \mp \Ji^{1/4}g(\hat{P}_\mathrm{p}) + \sqrt{\hat{P}_\mathrm{p} \Ji} - 1}{\sqrt{\hat{P}_\mathrm{p}(0)\Ji} - 1}\right]\right\}.
\label{eq:stress_solution_flat_profile}
\end{equation}
Finally, $\hat{P}_\mathrm{p}b$ is set by the top wall boundary condition, Eq.~\ref{eq:BC}, which makes Eq.~\ref{eq:stress_solution_flat_profile} an implicit but complete solution for the stress profile.


\end{document}